\documentclass{myegpubl}
\usepackage{sca2018}

\SpecialIssuePaper         
 \electronicVersion 

\ifpdf \usepackage[pdftex]{graphicx} \pdfcompresslevel=9
\else \usepackage[dvips]{graphicx} \fi

\PrintedOrElectronic

\usepackage{t1enc,dfadobe}

\usepackage{egweblnk}
\usepackage{cite}

\usepackage{siunitx}  
\usepackage{caption}
\usepackage{subcaption}
\usepackage{amsmath}
\usepackage{mathbbol}

\newcommand{\myparagraph}[1]{\noindent\textbf{#1}}
\graphicspath{{figs/}}

\title[Liquid Splash Modeling with Neural Networks]%
      {Liquid Splash Modeling with Neural Networks}

\author[Um et al.]
{\parbox{\textwidth}{\centering Kiwon Um,~
        Xiangyu Hu,~
        and Nils Thuerey
        }
        \\
{\parbox{\textwidth}{\centering Technical University of Munich, Germany
       }
}
}

\begin{document}

\teaser{
  \centering
  \includegraphics[width=0.33\linewidth]{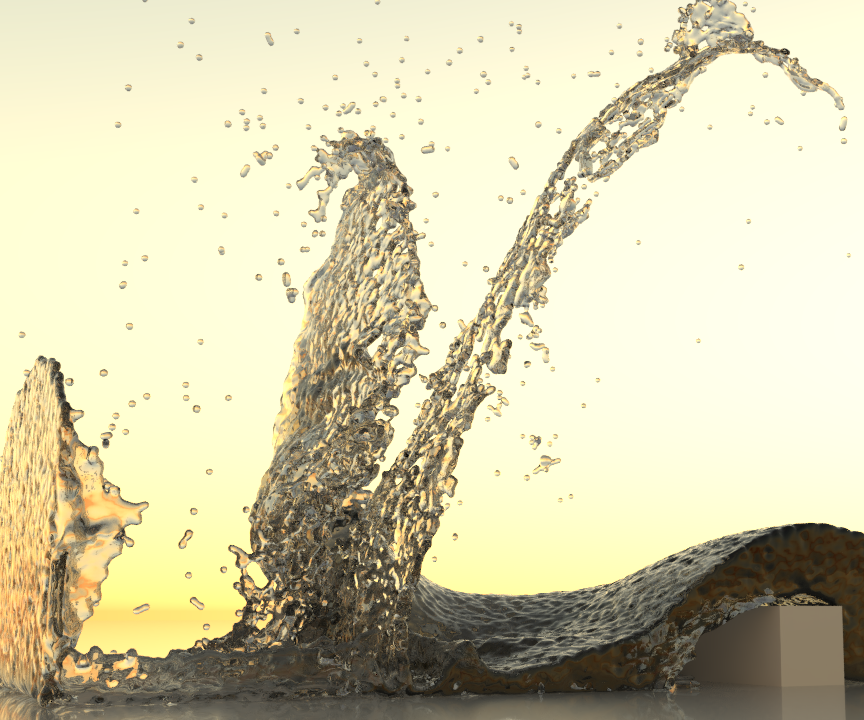}
  \includegraphics[width=0.33\linewidth]{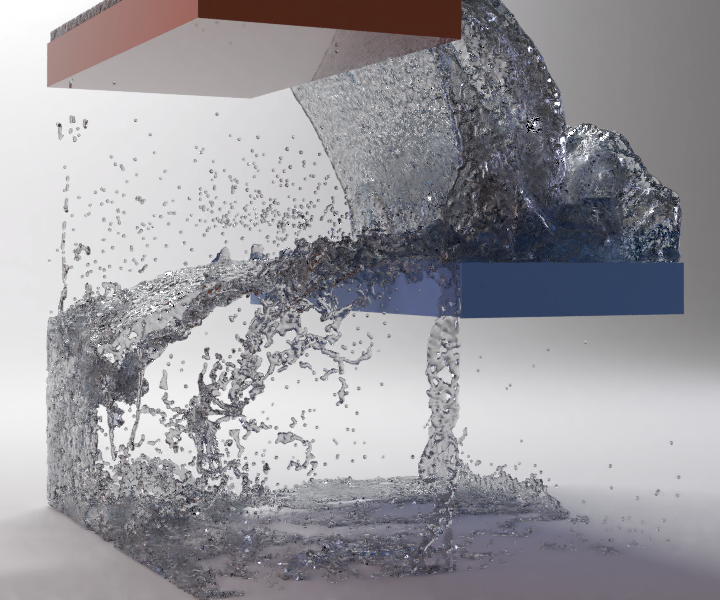}
  \includegraphics[width=0.33\linewidth]{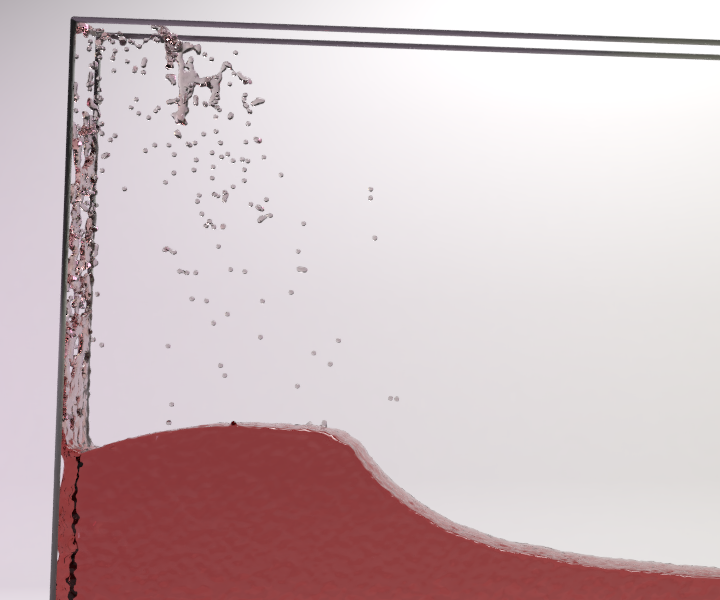}
  \caption{Our data-driven splash model improves the visual fidelity of FLIP
    simulations as shown here with three different simulation setups. Our model
    learns to infer the probability and velocity changes of under-resolved
    droplet formation effects with the trained neural networks.\\}
  \label{teaser}
}

\maketitle

\begin{abstract}
  This paper proposes a new data-driven approach to model detailed splashes for
  liquid simulations with neural networks. Our model learns to generate
  small-scale splash detail for the fluid-implicit-particle method using
  training data acquired from physically parametrized, high resolution
  simulations. We use neural networks to model the regression of splash
  formation using a classifier together with a velocity modifier. For the
  velocity modification, we employ a heteroscedastic model. We evaluate our
  method for different spatial scales, simulation setups, and solvers. Our
  simulation results demonstrate that our model significantly improves visual
  fidelity with a large amount of realistic droplet formation and yields splash
  detail much more efficiently than finer discretizations.\\

\ccsdesc[500]{Computing methodologies~Physical simulation}
\ccsdesc[300]{Computing methodologies~Supervised learning by regression}

\printccsdesc
\end{abstract}

\section{Introduction}
\label{sec:introduction}

For large-scale liquid simulations, it is crucial for visual fidelity that a
numerical simulation can produce a sufficient number of very small droplets
\cite{gerszewski2013,um2017percept}. However, it is difficult to capture such
splashes in practical simulations due to their complex small-scale surface
geometry and dynamics. A typical remedy for this difficulty is to use additional
representations that are physically or heuristically inspired to improve the
visual realism of the underlying simulations
\cite{takahashi2003,kim2006,ihmsen2012}. As an alternative, small drops can be
also tracked as part of the underlying simulation method \cite{guendelman2005}.
However, accurately resolving the effects of drop formation typically requires
the use of very fine spatial discretization, which in turn leads to high
computational cost. Thus, it is often challenging to generate vivid splashes in
liquid simulations as they require resolving the small-scale dispersive motions
that lead to droplets forming and being ejected from the bulk volume.

This paper proposes a new data-driven splash generation approach that improves
the visual fidelity of hybrid particle-grid liquid simulations. By learning the
formation of small-scale splashes from physically accurate simulations, our
model effectively approximates the sub-grid scale effects that lead to droplet
generations. This enables us to generate realistic splashes in coarse
simulations without the need for manually tweaking parameters or increased
computational costs induced by high resolution simulations. We realize our model
using machine learning techniques with neural networks (NNs) and integrate the
model into the fluid-implicit-particle (FLIP) algorithm \cite{zhu2005}. Within
this environment, we investigate the generality of our design by employing
different simulation methods for generating the training data. Moreover, we show
an extension of our model to generate secondary particles, which are independent
from the underlying simulation. Using this extension, we also investigate
controlling the number of generated splashes and the ability of our model to
learn from multi-scale data. Figure~\ref{teaser} shows three examples of results
that our model can generate. In the following, we will refer to our model as
\emph{MLFLIP}, which indicates a combination of machine learning and FLIP.

\section{Related Work}

The behavior of liquids is typically modeled with the \emph{Navier-Stokes} (NS)
equations:
\begin{equation}
  \label{eq:ns}
  \frac{\partial\mathbf{u}}{\partial{t}} + \mathbf{u} \cdot \nabla \mathbf{u} =
  \mathbf{g} - \frac{1}{\rho}\nabla{P} + \nu\nabla^2\mathbf{u}
  \quad\mathrm{and}\quad \nabla\cdot\mathbf{u} = 0,
\end{equation}
where $\mathbf{u}$ is the velocity, $\mathbf{g}$ is the gravity, $\rho$ is the
density, $P$ is the pressure, and $\nu$ is the viscosity coefficient. There
exist many numerical methods for solving these equations, which are commonly
categorized as Eulerian and Lagrangian approaches
\cite{bridson2015book,ihmsen2014star}.

FLIP is a particularly popular method for liquid simulations \cite{zhu2005}, and
it is widely used in movie visual effects at the moment. Its effective
combination of Lagrangian and Eulerian properties enables the efficient solve of
liquid motions. While FLIP has become a practical solution for liquid
simulations, the core method has been extended to various ways in order to
improve its simulation quality and efficiency. For example, different position
correction methods improved the distribution of particles
\cite{ando2012,um2014}, and Ferstl \textit{et al.} \shortcite{ferstl2016}
proposed a narrow band method that improves the efficiency by sampling the
volume with particles only near the surface. In addition, FLIP has been widely
adopted to various problems of fluid simulations such as viscous free surfaces
\cite{batty2008} and solid-fluid coupling \cite{batty2007}. Two-phase fluids
have been also simulated via an extension of FLIP \cite{boyd2012}. Moreover, the
affine particle-in-cell (APIC) method \cite{jiang2015} as a variant effectively
addressed the stability issues of FLIP and the dissipation of the
particle-in-cell method, and APIC has been further generalized to polynomial
representations \cite{fu2017}.

A central goal of our model is to improve the visual fidelity of liquid
simulations with small-scale details. Apart from FLIP, the smoothed particle
hydrodynamics (SPH) approach is a popular alternative in graphics
\cite{muller2003,solenthaler2009,akinci2013c,bender2017}. Due to the nature of
its computational mechanism based on interparticle interactions, SPH is able to
simulate dispersive motions and droplets. This behavior led to a combination of
SPH and the particle level set approach \cite{enright2002}; this simulates the
diffuse regions via SPH \cite{losasso2008}. Our model infers such small-scale
interactions in a data-driven way. In addition, we employ an efficient hybrid
particle-grid solver (i.e., FLIP). Thus, our method does not require potentially
expensive particle neighborhood information. A possible alternative approach to
achieve this goal in FLIP is to add details using extra representations for
diffuse materials \cite{yang2014,yang2015}. Likewise, Ihmsen \textit{et al.}
\shortcite{ihmsen2012} proposed a flexible secondary particle model for such
diffuse materials in SPH simulations. With enough manual tuning and elaborate
coupling processes, these extra representation approaches can yield realistic
results, but in contrast to their work, we focus on an automated approach that
captures splash effects for physically-parametrized real world scales. Our model
does not require any parameter tuning on the user side and sophisticated
integration. At the same time, one of our goals is to demonstrate that NNs are
suitable to detect and generate these splashes.

A method that shares our goal to enable splashes with FLIP is the unilateral
incompressibility (UIC) solver \cite{gerszewski2013}. The UIC solver allows
positive divergence in fluid cells such that it can create larger-scale
splashes. However, it leads to a very different visual behavior as the grid
based velocities lead to a formation of sheets and filaments rather than
detaching droplets. In addition, the UIC approach requires two solves of the
linear complementarity problem. Our approach targets a very different direction.
Instead of modifying the pressure solve, we incorporate a statistical model with
the help of machine learning. We note that our approach is orthogonal to the
choice of Poisson solver and thus could also integrate into the UIC solver in a
single simulation to increase the small-scale splash effects.

\myparagraph{Machine learning:} Machine learning is a field that recently
received substantial attention due to the success of so-called deep neural
networks. Here, especially the seminal image classification work of Krizhevsky
\textit{et al.} \shortcite{krizhevsky2012} has to be mentioned, but other areas
such as object detection \cite{girshick2014} and control of complex systems
based on visual inputs \cite{mnih2013} have likewise seen impressive advances.
As our splash model utilizes NNs, we briefly review their concepts and give an
overview of previous work on NNs for physics simulations. In general, the
learning process aims for the approximation of a general function $f$ using a
given data set (i.e., input $\mathbf{x}$ and output $\mathbf{y}$) in the form of
$\mathbf{y} = f(\mathbf{x}, \mathbf{w})$ where $\mathbf{w}$ is the set of
weights and biases to be trained. Using NNs, the general function $f$ is modeled
by networks of multiple layers where each layer contains multiple nodes (i.e.,
artificial neurons). These networks consist of layers with connected nodes. The
output vector $\mathbf{y}_L$ from a layer $L$ is typically computed with
$\mathbf{y}_{L} = \Phi(\mathbf{w}_{L} \mathbf{y}_{L-1} + \mathbf{b}_{L})$ where
$\Phi(\cdot)$ is the activation function that is applied to each component,
$\mathbf{w}_{L}$ is the weight matrix of the layer, and $\mathbf{b}_{L}$ is the
bias vector of the layer. Here, the activation function $\Phi$ enables the
networks to capture non-linearities in the approximated function. We will
demonstrate that NNs, which so far have rarely been used for fluid simulations,
can be used for realization of our data-driven splash model.

NNs were previously used to compute local pressure approximations
\cite{yang2016} while others employed networks with convolutional layers to
regress the whole pressure field \cite{tompson2017}. Moreover, an approach using
regression forests, which represent an alternative to NNs, was proposed to
efficiently compute forces in SPH \cite{ladicky2015}. More recently, NNs were
also successfully employed for patch-based smoke simulations \cite{chu2017},
super-resolution with temporal coherence \cite{xie2018}, and fast generation of
liquid animations with space-time deformations \cite{rtliquids2017}. In the
engineering community, approximating effects smaller than the discretization
resolution is known as \emph{coarse-grained modeling} \cite{hijon2009}, but this
idea has not been used to model splash formation. We propose to use machine
learning techniques to represent accurate and high resolution data in order to
approximate complex small-scale effects with high efficiency.

\section{Data-driven Splash Modeling}
\label{sec:nns}

The following section details our data-driven approach for generating splashes.
The principal idea of our approach is to infer statistics about splash formation
based on data from simulations that are parametrized to capture the droplet
formation in nature. Our definition of a \emph{splash} is a small disconnected
region of liquid material that is not globally coupled with the main liquid
volume. The key novelty of our approach is that it does not require manually
chosen parameters, such as velocity or curvature thresholds, to generate the
splashes. Rather, we use a statistical model and data extracted from a series of
highly detailed and pre-computed simulations.

Our approach consists of two components: a \emph{detachment classification} and
a \emph{velocity modification}. Based on a feature descriptor consisting of
localized flow information, the classifier predicts whether a certain volume of
liquid forms a detached droplet within a chosen duration $\Delta{t}$ (typically,
on the order of a single frame of animation). For droplets that are classified
as such, our modifier predicts its future velocity based on a probability
distribution for velocity modifications. We use NNs to represent both
components, and the following sections describe our statistical model and the
corresponding machine learning approach.

\subsection{Neural Network Model}
\label{sec:nns-nets}

\begin{figure}[tb]
  \centering
  \def\svgwidth{0.9\linewidth}
  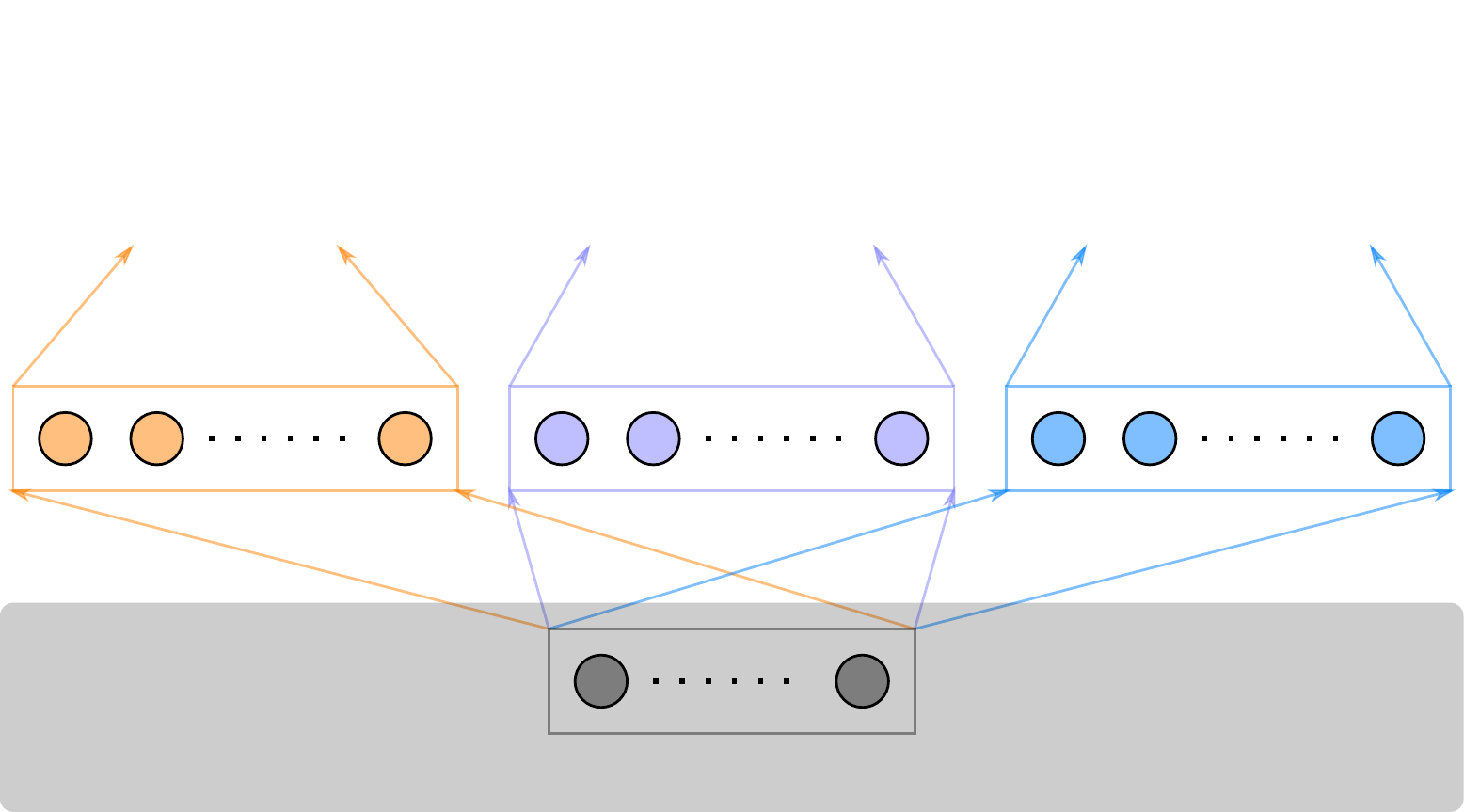
  \caption{The overall structure of our neural networks.}
  \label{fig:nns}
\end{figure}

The input to our model is a feature descriptor $\mathbf{x} \in \mathbb{R}^M$
that encapsulates enough information of the flow such that a machine learning
model can infer whether the region in question will turn into a splash. For this
binary decision ``\emph{splash} or \emph{non-splash}'', we will use an indicator
value $l \in \{1, 0\}$ in the following. Each descriptor is typically generated
for a position $\mathbf{p}$. The $M$ individual components of $\mathbf{x}$
consist of flow motion and geometry in the vicinity of $\mathbf{p}$. In
practice, we use $3^3$ samples of the velocity and level set. The discussion of
this choice will be given in Section~\ref{sec:nns-data} in more detail.

We train our models with a given data set that consists of feature vectors
$\mathbf{X} = \{\mathbf{x}_1, \mathbf{x}_2, \cdots, \mathbf{x}_N\}$ and
corresponding detachment indicator values
$\mathbf{L} = \{l_1, l_2, \cdots, l_N\}$; they are generated during a
pre-processing phase at locations
$\{\mathbf{p}_1, \mathbf{p}_2, \cdots, \mathbf{p}_N\}$. Then, our classifier
aims for inferring the probability $P_s$ that a feature vector $\mathbf{x}_i$ is
in the class indicated by $l_i$. Considering a probability distribution function
$\mathbf{y}_s$ that $P_s$ follows, we will approximate the function
$\mathbf{y}_s$ from the given data. For this task, we can follow established
procedures from the machine learning field \cite{bishop2006}.

The probability distribution $\mathbf{y}_s(\mathbf{x}_i, \mathbf{w}_s)$ is the
target function that is represented by the weights $\mathbf{w}_s$. The weights
are the actual degrees of freedom that will be adjusted in accordance to the
data during the learning phase. We can express $P_s$ in terms of $\mathbf{y}_s$
as:
\begin{equation}
  \label{eq:p_s}
  P_s(l_i|\mathbf{x}_i) \sim P\left(l_i|\mathbf{y}_s(\mathbf{x}_i, \mathbf{w}_s)\right),
\end{equation}
which yields the following likelihood function that we wish to maximize:
\begin{equation}
  \label{eq:L_d}
  L_d(\mathbf{L}|\mathbf{X}) =
  \prod_{i=1}^N P\left(l_i | \mathbf{y}_s(\mathbf{x}_i, \mathbf{w}_s)\right).
\end{equation}
In order to maximize this likelihood, we use the well-established \emph{softmax}
(i.e., normalized exponential function) for the loss of our classification
networks. It will successfully encode the given data set and train the model for
$\mathbf{y}_s$, and then we can evaluate with new feature vectors at any
position in a flow to predict whether the region will turn into a detached
droplet within the time frame $\Delta{t}$.

\begin{figure*}[t]
  \centering
  \includegraphics[width=\linewidth]{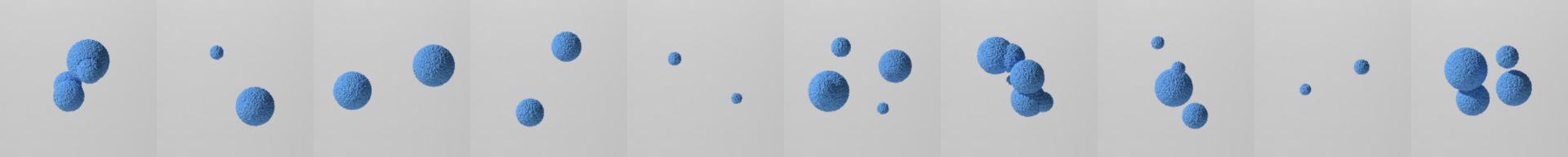}
  \caption{Selected close-up snapshots of ten randomized initial conditions to
    generate the droplet formation data for training.}
  \label{fig:training-sim-init}
\end{figure*}
\begin{figure*}[t]
  \centering
  \includegraphics[width=\linewidth]{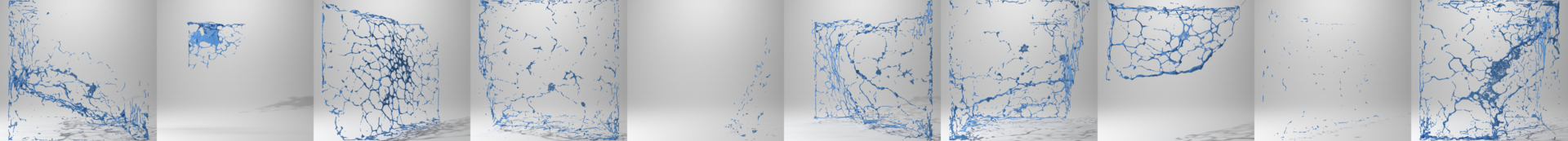}
  \includegraphics[width=\linewidth]{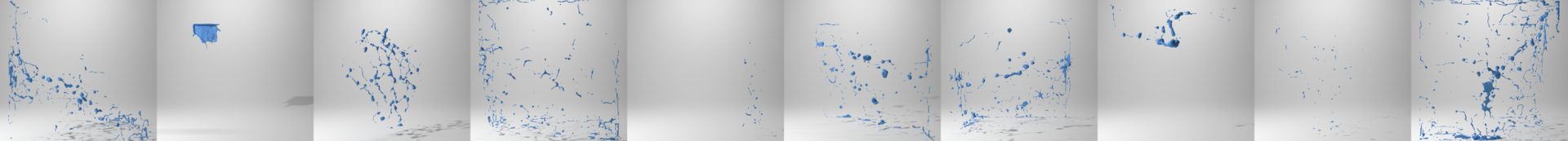}
  \caption{Example frames of (top) 5\si{mm} and (bottom) 1.5\si{mm} scale
    simulations for training data generation.}
  \label{fig:training-sim-frame}
\end{figure*}

Let $\Delta{\mathbf{v}}$ be an instance of a velocity change for a splash with
respect to the motion of the bulk liquid in its vicinity. We will afterward
consider this velocity change of a droplet relative to the bulk as the
\emph{velocity modification} of a particle in our simulation. Similar to the
classifier above, our goal is to infer the set of velocity modifications
$\Delta{\mathbf{V}} = \{\Delta{\mathbf{v}_1}, \Delta{\mathbf{v}_2}, \cdots,
\Delta{\mathbf{v}_N}\}$ based on the corresponding set of feature vectors
$\mathbf{X}$. From the statistics of our training data, we found that it is
reasonable to assume that the velocity modifications follow a normal
distribution relative to the mean flow direction. Accordingly, we model the
modifier as a modification function $f_m(\Delta{\mathbf{v}_i}|\mathbf{x}_i)$
following a normal distribution with mean $\boldsymbol{\mu}$ and variance
$\boldsymbol{\sigma}^2$:
\begin{equation}
  \label{eq:f_m}
  f_m(\Delta{\mathbf{v}_i}|\mathbf{x}_i) \sim
  \mathcal{N}\left(
  \Delta{\mathbf{v}_i}|{\boldsymbol{\mu}}(\mathbf{x}_i, \mathbf{w}_{\boldsymbol{\mu}}),
  {\boldsymbol{\sigma}^2}(\mathbf{x}_i, \mathbf{w}_{\boldsymbol{\sigma}^2})\right),
\end{equation}
thus
\begin{equation}
  \label{eq:f_m_N}
  f_m(\Delta{\mathbf{v}_i}|\mathbf{x}_i) \sim
  \frac{1}
  {\sqrt{2\pi{\boldsymbol{\sigma}^2_i}}}
  \exp\left(
    {\frac{-(\Delta{\mathbf{v}_i} - {{\boldsymbol{\mu}_i}})^2}{2{\boldsymbol{\sigma}^2_i}}}
  \right),
\end{equation}
where, for the sake of simplicity, ${\boldsymbol{\mu}_i}$ and
${\boldsymbol{\sigma}^2_i}$ denote
${\boldsymbol{\mu}}(\mathbf{x}_i, \mathbf{w}_{\boldsymbol{\mu}})$ and
${\boldsymbol{\sigma}^2}(\mathbf{x}_i, \mathbf{w}_{\boldsymbol{\sigma}^2})$,
respectively. Then, the negative log likelihood function $L_m$ (also known as
loss function) for the given data is defined as follows:
\begin{equation}
  \label{eq:loss-mve}
  L_m(\Delta{\mathbf{V}}|\mathbf{X}) = \frac{1}{2} \sum_{i=1}^N \sum_{j=1}^d \left[
    \frac{(\Delta{\mathbf{v}_{i,j}} - {{\boldsymbol{\mu}}_{i,j}})^2}{{\boldsymbol{\sigma}^2_{i,j}}}
    + \ln{{\boldsymbol{\sigma}^2_{i,j}}}
  \right],
\end{equation}
where $j$ denotes the spatial index.

As Eq.~(\ref{eq:loss-mve}) indicates, we model the modifier as a mean variance
estimation (MVE) problem \cite{nix1994,khosravi2011}. Instead of directly
estimating the mean of targets, the MVE problem assumes that errors are normally
distributed around the mean and estimates both the mean and
\emph{heteroscedastic} variance. Note that
${\boldsymbol{\mu}}(\mathbf{x}_i, \mathbf{w}_{\boldsymbol{\mu}})$ and
${\boldsymbol{\sigma}^2}(\mathbf{x}_i, \mathbf{w}_{\boldsymbol{\sigma}^2})$ are
the target functions that are approximated with each set of weights
$\mathbf{w}_{\boldsymbol{\mu}}$ and $\mathbf{w}_{\boldsymbol{\sigma}^2}$.

In our approach, the mean and variance functions are represented by NNs and
approximated by estimating the two sets of weights such that they minimize the
loss function $L_m$ for the given data $\{\mathbf{X}, \Delta{\mathbf{V}}\}$. We
want to point out that several machine learning algorithms are available to
solve this problem \cite{bishop2006}, but NNs have proven themselves to be
robust for this problem, and we found that they work very well in our tests.

The overall structure of our NNs is illustrated in Figure~\ref{fig:nns}. The NNs
learn for two separate components: classifier and modifier. Sharing the input
vector $\mathbf{x}$, both components are represented as separate two-layer NNs.
The size of output from the first layer is double the input vector's, and the
output is fully connected to the final output. All outputs from each layer are
activated using the hyperbolic tangent function. Note that there is a large
variety of different network layouts that could tried here, but we found that
this simple structure worked sufficiently well in our tests.

\begin{figure*}[tb]
  \centering
  \includegraphics[width=\linewidth]{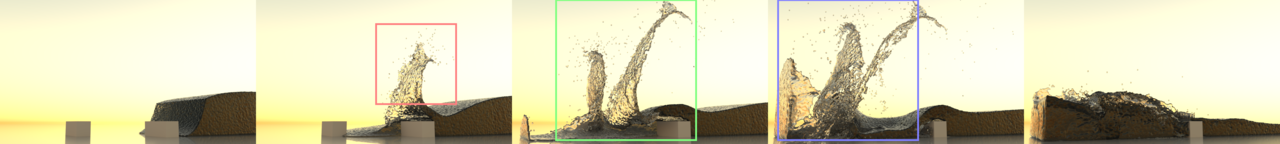}
  \includegraphics[width=0.33\linewidth]{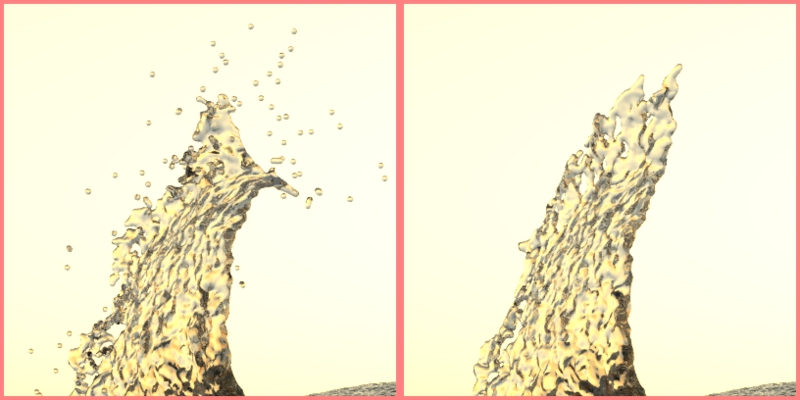}
  \includegraphics[width=0.33\linewidth]{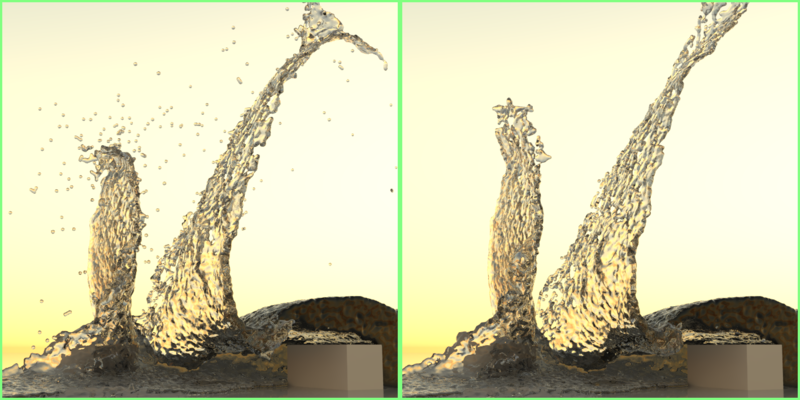}
  \includegraphics[width=0.33\linewidth]{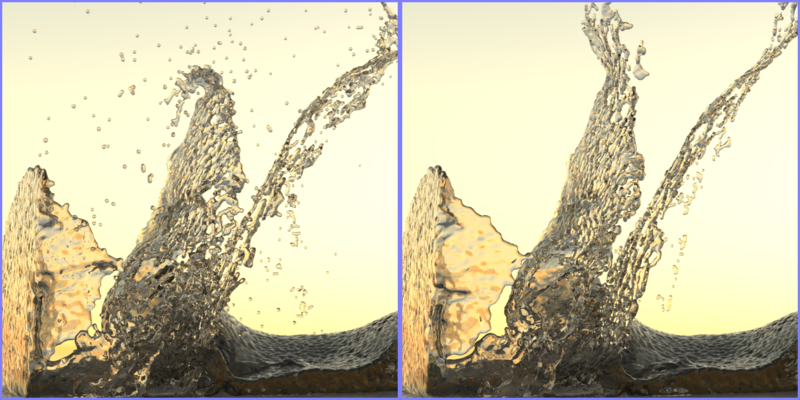}
  \caption{Example frames of dam simulations with MLFLIP. The bottom row
    compares (left) MLFLIP with (right) FLIP side-by-side in the selected areas
    of three frames.}
  \label{fig:dam}
\end{figure*}

\subsection{FLIP Integration}
\label{sec:nns-flip}

Our NN-based splash generation model easily integrates into an existing FLIP
simulation pipeline. After the pressure projection step, we run classification
on all particles in a narrow band of one cell around the surface. For those that
generate a positive outcome, we evaluate our velocity modification network to
compute component-wise mean and variance. Then, we generate random numbers from
correspondingly parametrized normal distributions to update the velocity of the
new splash particle. All splashes are treated as ballistic particles and do not
participate in the FLIP simulation until they impact the main volume again.
Thus, we treat individual splashing droplets as particles that only experience
gravitational acceleration but no other NS forces. This modeling is in line with
the secondary effects often employed in movies \cite{losure2012}.

Our model evaluates the probability of forming a splash in regard to a chosen
duration $\Delta t$, which is typically larger than the simulation time-step
size because of the stability constraint. Unless this difference is carefully
considered, the time-step size will affect the inference results. As we use
smaller time-step size so evaluate more frequently, we will get more splashes.
To avoid this, we evaluate our inference using a stochastic process
\cite{papoulis2002}, i.e., random walk, scaling the expectation by time to match
the desired duration. For a series of simulation steps $\Delta t_k$, which
proceed for the chosen duration $\Delta t$ (i.e.,
$\sum_k \Delta t_k = \Delta t$), we compute particle $i$'s expectation $E_i$
using two finite outcomes 1 (splash) and -1 (non-splash) with probabilities
$\mathbf{y}_s$ that are evaluated from our trained model:
\begin{equation}
  \label{eq:expectation}
  E_i = \sum_k \frac{1}{\sqrt{{\Delta t}/{\Delta t_k}}} \mathbf{y}_{s}(\mathbf{x}_{i,k}, \mathbf{w}_s) \cdot [1, -1].
\end{equation}
If the expectations of particles are positive after the random walks during
$\Delta t$, we treat them as splashes and thus evaluate our velocity
modification.

FLIP can use different numbers of simulation particles for the same grid
resolution, and this will affect the number of inferred splashes since our model
evaluates splashes for each simulation particle. In order to make the model
robust to such a variety, we can normalize the inference process in space. While
the expectations are evaluated for each particle, we can limit the maximum
number of splashes per unit volume using the expected values $E_i$ (e.g., one
particle whose expectation is the largest value in a cell). We found that our
approach generates a consistent number of splashes regardless of the size of the
time-step and the number of simulation particles per grid cell.

\myparagraph{Look-ahead correction:} While our splash generation algorithm
reliably works in our tests, we noticed a small chance of misclassification.
This can happen, for instance, when the side of an obstacle is just outside the
region of our input feature vector. To minimize the influence of such
misclassifications, we implemented a look-ahead step that reverts the
classification of individual splashes if the droplets do not manage to form the
expected splashes. For this look-ahead check, we advance all bulk volume
particles, i.e., those that were not classified as splashes, by $\Delta t$ using
the current grid velocities. We separately integrate positions of the new splash
particles for a time interval of $\Delta t$ with their updated velocities. If a
new splash particle ends up inside of the bulk liquid or an obstacle, we revert
its classification and modification.

\subsection{Training Data}
\label{sec:nns-data}

Our NNs require a large set of input feature vectors with target outputs for
training. In our model, the latter consists of the classification result $l$
indicating whether a flow region forms a splash and the velocity modification
$\Delta{\mathbf{v}}$ predicting the trajectory of a splash. We generate the
training data from a series of simulations with randomized initial conditions
designed to incite droplet formation. The randomized initial conditions are the
number of droplets and their initial positions and velocities. We choose the
ranges of each condition such that they yield sufficient variability for data
generation. The snapshots of the randomized example simulations are shown in
Figure~\ref{fig:training-sim-init}. Note that at this stage any available
``accurate'' NS solver could be employed. However, we demonstrate that FLIP can
\emph{bootstrap} itself by using high resolution simulations with correctly
parametrized surface tension.

For the training data simulations, it is crucial that they resolve important
sub-grid effects that are not well resolved on coarse resolutions to which our
model will be applied later. We test our model with two physical
parametrizations where the surface tension is dominant, and thus many droplets
are generated. Our training simulations are performed using FLIP with the ghost
fluid method \cite{hong2005} for surface tension effects. The two scales use a
grid spacing of 5\si{mm} and 1.5\si{mm}, and they are parametrized with the
surface tension of water, i.e., 0.073 \si{N/m}. Both scales use a simulation
grid of 100$^3$. Several example frames for both scales are shown in
Figure~\ref{fig:training-sim-frame}.

Note that the simulation data for training will be used to encode the desired
sub-grid effects for much larger scales afterwards. When applying our model in
new simulation setups, we typically have scales that are much coarser than those
used for generating the training data. Thus, the feature descriptor (i.e.,
$\mathbf{x}_i$) needs to be defined at this coarse simulation scale, and our
networks need to infer their outputs (i.e., $l_i$ and $\Delta{\textbf{v}}_i$)
based only on this coarse input. For this purpose, we make use of a coarse grid
at data generation time. This coarse grid represents the scale to which the
model will be applied afterward. For every time step, we down-sample the
necessary high resolution fields from the data generation simulation to this
lower resolution and extract the feature descriptors for training.

As indicated in Section~\ref{sec:nns-nets}, we define a feature descriptor
$\mathbf{x}_i$ using $3^3$ samples of the velocity and level set values
interpolated with a sampling spacing of $h$; i.e., the feature vector consists
of 108 components containing $3^3 \times 3$ velocity values and $3^3 \times 1$
level set values. Because the splash formation mostly relies on the local flow
physics near the liquid surface, we focus on the localized flow information and
the surface region where the splash is likely initiated. From pilot experiments,
we found that the improvement is negligible when more samples or more features
such as obstacle information are used for the feature vector.

\begin{figure*}[tb]
  \centering
  \captionsetup[subfigure]{aboveskip=0em,belowskip=0em}
  \begin{subfigure}[b]{0.69\linewidth}
    \includegraphics[width=\linewidth]{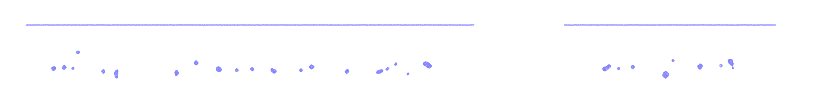}
    \includegraphics[width=\linewidth]{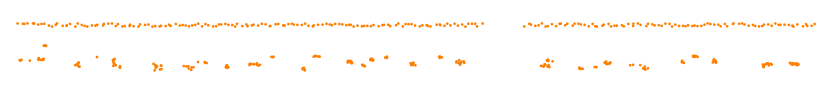}
    \caption{}
    \label{fig:string:ex}
  \end{subfigure}
  \begin{subfigure}[b]{0.29\linewidth}
    \includegraphics[width=\linewidth]{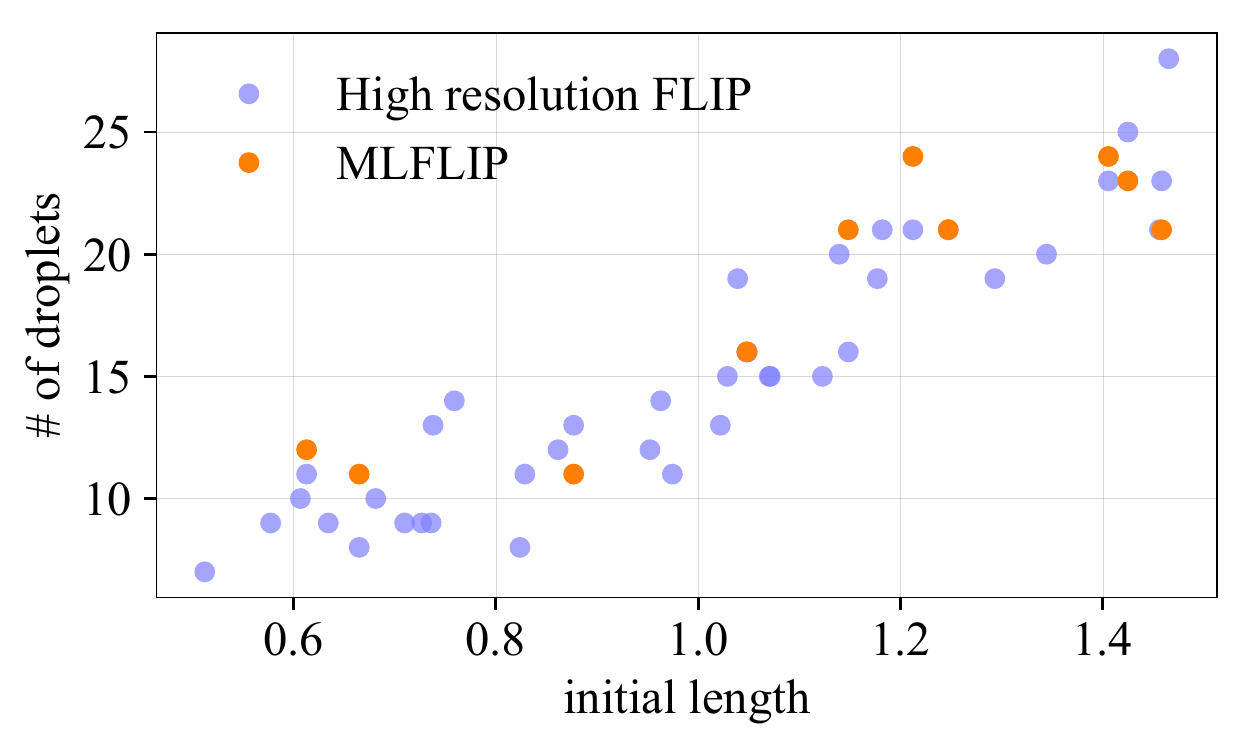}
    \caption{}
    \label{fig:string:graph}
  \end{subfigure}
  \caption{Number of droplets formed after 1.5 seconds with different initial
    lengths. (a) Two selected high resolution FLIP simulations (top row in blue)
    show the surface tension driven dynamics. Two examples of our MLFLIP
    simulations (bottom row in orange) successfully recover the break-up in a
    low resolution simulation. Droplet count statistics are shown in (b). Our
    simulations (orange dots) very accurately match the full resolution
    simulation results (blue dots).}
  \label{fig:strings}
\end{figure*}

In order to extract the splash indicator value $l$, we analyze the particle's
spatial motion for a chosen duration $\Delta{t}$ (i.e., a single frame of
animation in our experiments). Using an auxiliary grid, the separate volumes are
recognized by computing the isolated liquid regions from the level set field or
cell flags. We then identify the splashing particles (i.e., $l$=1) as those
ending up in a new disconnected component that falls below a given volume
threshold at time $t+\Delta{t}$. In our case, if a disconnected component
consists of less than 8 cells, the volume is marked as droplet volume. All
particles in such a droplet are marked as splashes if the droplet did not exist
as a disconnected component at time $t$.

The velocity modifier of our model predicts the trajectory for a splash. We
evaluate this prediction after updating the velocity of particle from the
divergence-free velocity at the target resolution. Thus, the new velocity
$\mathbf{v}_m^{t+\Delta{t}}$ for a splash particle is defined as
$ \mathbf{v}_m^{t+\Delta{t}} = \mathbf{v}^{t+\Delta{t}} + \Delta{\mathbf{v}}$,
and we compute the velocity modification $\Delta{\mathbf{v}}$ for training:
\begin{equation}
  \label{eq:dv}
  \Delta{\mathbf{v}} = 
  \frac{\mathbf{p}^{t+\Delta{t}} - \mathbf{p}^{t}}{\Delta{t}} - \mathbf{v}^{t+\Delta{t}}
\end{equation}
where, intuitively, the first term on the right side represents the velocity of
the training resolution, and the second term represents the down-sampled
velocity evaluated at the target resolution.

As the splashes are initiated near the liquid surface in general, we extract the
training data only from the surface particles. The surface particles are
recognized by slightly expanding the area of empty (i.e., air) cells. Note that
we use the data from splashing as well as non-splashing particles for training.
It is crucial for training that the networks see sufficiently large numbers of
samples from both classes.

For training, we used 1.3M samples of the 5\si{mm} scale and 441K samples of the
1.5\si{mm} scale. They were extracted from sixteen training simulations per
scale. The data of each scale contain the same number of both splashing and
non-splashing samples. Note that the same set of initial conditions
(Figure~\ref{fig:training-sim-init}) generates different physics in different
scales (Figure~\ref{fig:training-sim-frame}). The smaller scale incites less
splashes because the surface tension forces are more prominent. This resulted in
a reduced number of training samples for the smaller scale of 1.5\si{mm}.

\subsection{Training Networks}
\label{sec:training-nets}

\begin{figure}[b]
  \centering
  \includegraphics[width=.49\linewidth,page=1]{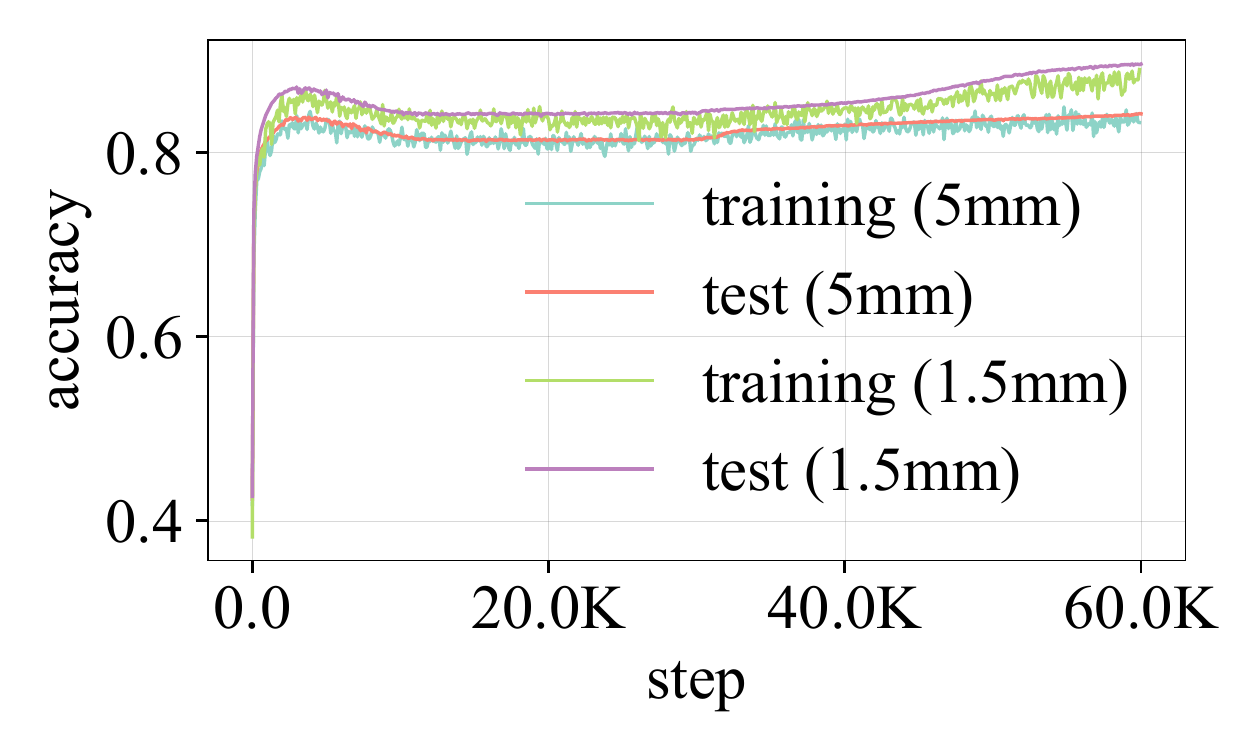}
  \includegraphics[width=.49\linewidth,page=2]{figs/tf-learning-curve}
  \caption{Learning graphs for both the training and test sets in two scales.
    The left graph shows the classification accuracy; the right graph shows the
    loss $L_m$.}
  \label{fig:learning-curves}
\end{figure}

We randomly split the training samples into 75\% for the \emph{training set} and
25\% for the \emph{test set}. The graphs in Figure~\ref{fig:learning-curves}
illustrate the progress of the learning phase in our experiments. The training
performed for 60K iterations with 5K samples as a training batch. When the data
sets were fully used, we randomly shuffled the samples and then continued with
the training. In order to train our MVE model, the first 30K iterations trained
the mean, keeping the variance constant, afterwards the remaining 30K iterations
trained both the mean and variance simultaneously \cite{nix1994}. Thus, as shown
in the right of Figure~\ref{fig:learning-curves}, we could observe two learning
phases: the first intermediate convergence of the mean value after ca. 15K
iterations and the final convergence approximately after 40K iterations.

To realize our NNs, we employed the \emph{TensorFlow} framework
\shortcite{tensorflow2015short} with its ADAM optimizer \cite{kingma2014}. Here,
the learning rate for training was set to $10^{-4}$. We also employed weight
decay and dropout (both with strength $10^{-1}$) to avoid over-fitting.
Additionally, we found that the batch normalization technique \cite{ioffe2015}
significantly improves the learning rate and accuracy.

\begin{figure*}[tb]
  \centering
  \includegraphics[width=\linewidth]{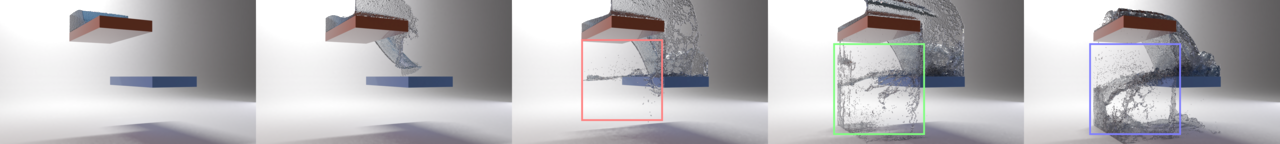}
  \includegraphics[width=0.33\linewidth]{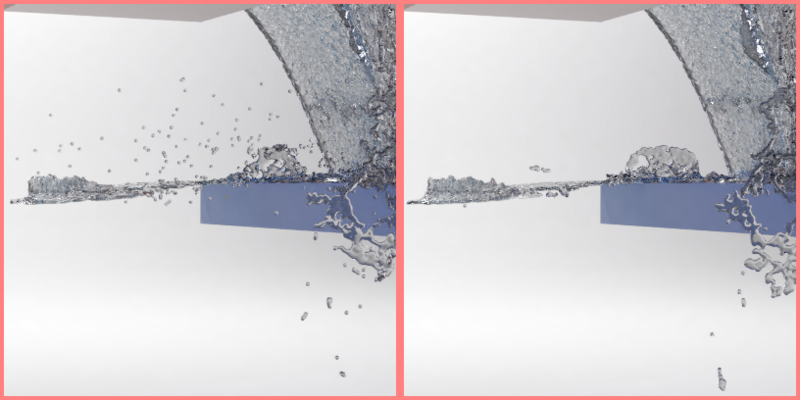}
  \includegraphics[width=0.33\linewidth]{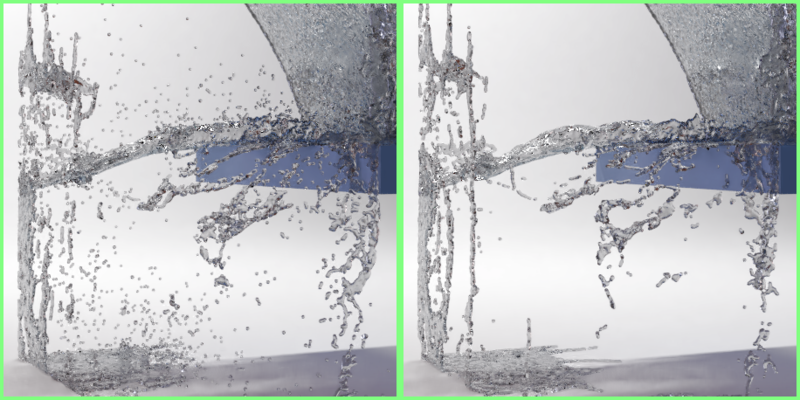}
  \includegraphics[width=0.33\linewidth]{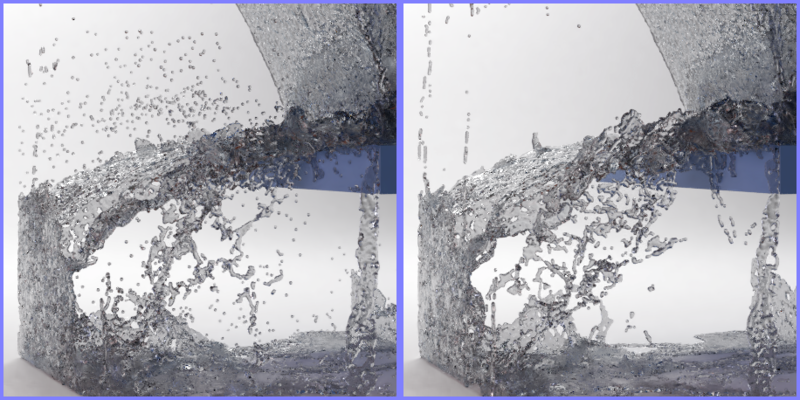}
  \caption{Example frames of stairs simulations with MLFLIP. The bottom row
    compares (left) MLFLIP with (right) FLIP side-by-side in the selected areas
    of three frames.}
  \label{fig:fall}
\end{figure*}

\section{Model Evaluation}
\label{sec:eval}

\begin{figure}[tb]
  \centering
  \captionsetup[subfigure]{aboveskip=0em,belowskip=0em}
  \begin{subfigure}[b]{\linewidth}
    \centering
    \includegraphics[width=\linewidth]{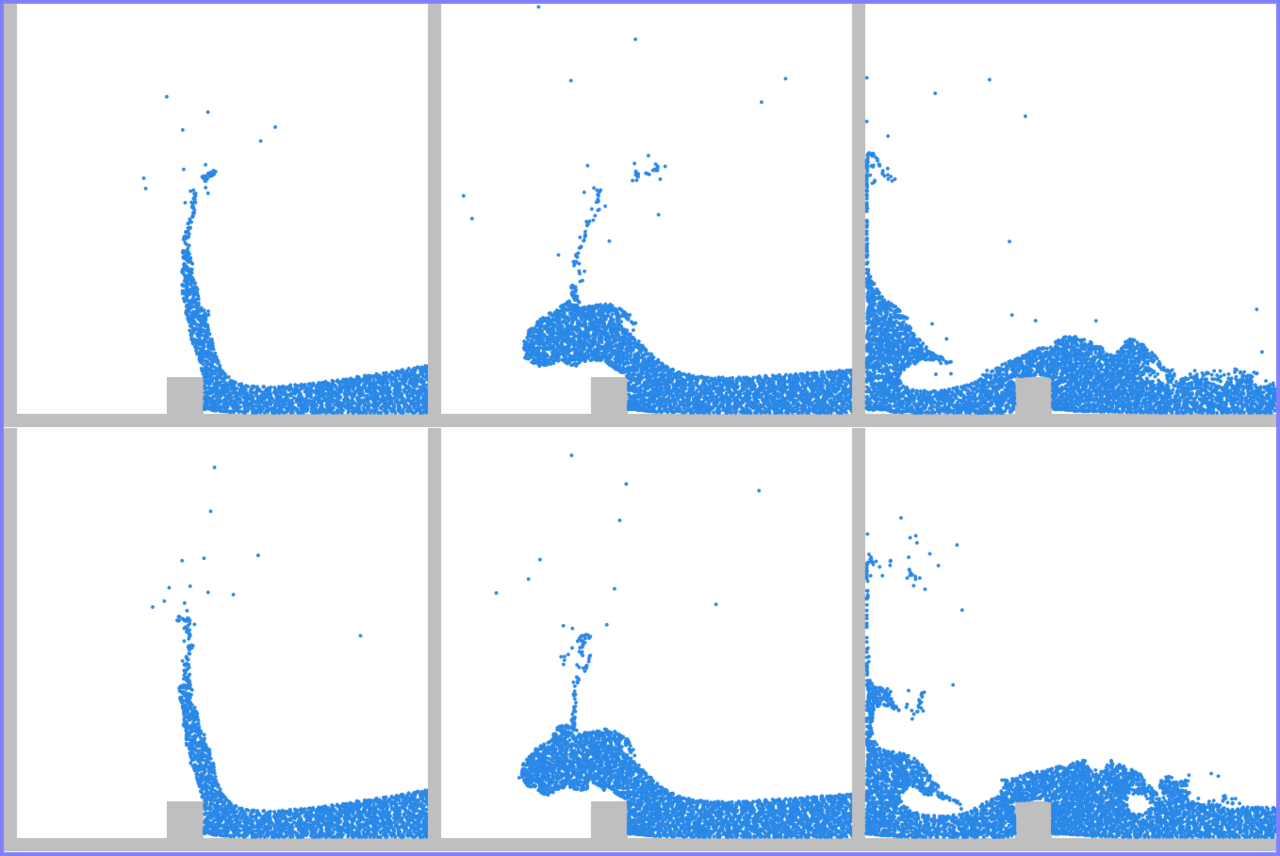}
    \caption*{\hfill{}MLFLIP}
  \end{subfigure}
  \begin{subfigure}[b]{\linewidth}
    \includegraphics[width=\linewidth]{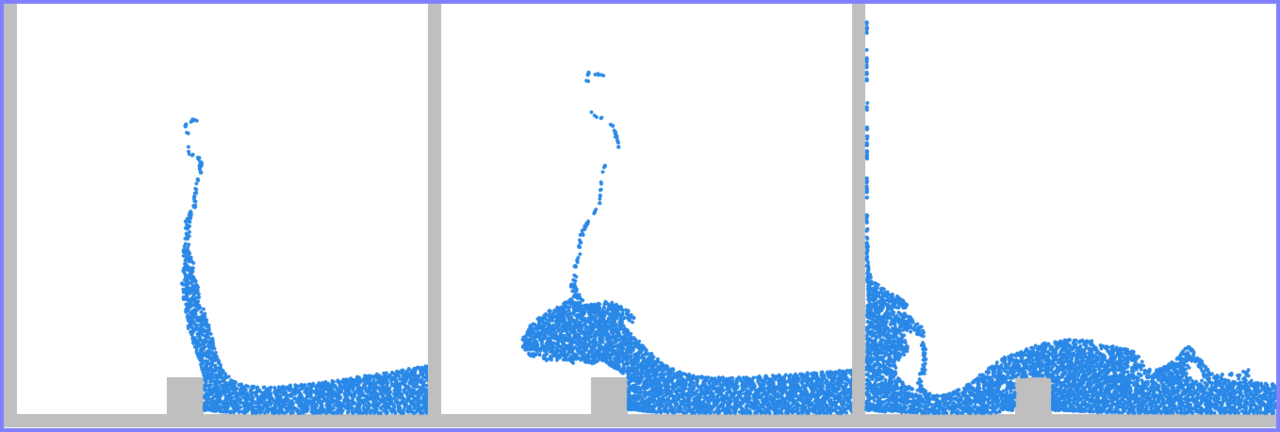}
    \caption*{\hfill{}FLIP}
  \end{subfigure}
  \caption{Comparison among (top) MLFLIP trained using FLIP, (middle) MLFLIP
    trained using SPH, and (bottom) FLIP.}
  \label{fig:mlflip-sph}
\end{figure}

\begin{figure}[tb]
  \centering
  \includegraphics[width=0.95\linewidth]{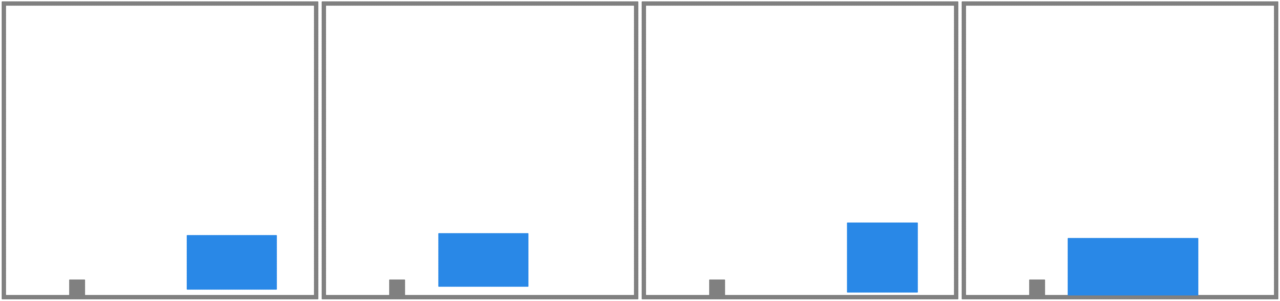}
  \includegraphics[width=0.95\linewidth]{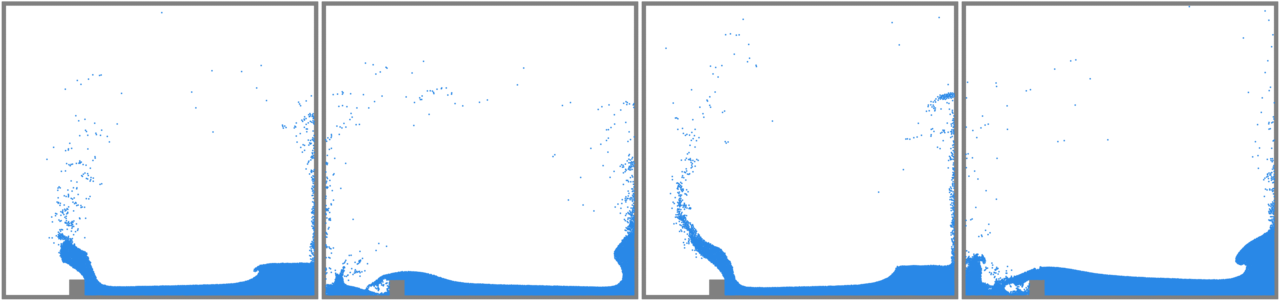}
  \caption{Example frames of SPH simulations for training data generation. The
    top row shows four selected initial conditions; the bottom row shows their
    snapshots after one second.}
  \label{fig:sph-data}
\end{figure}

\begin{figure}[tb]
  \captionsetup[subfigure]{aboveskip=0em,belowskip=0.1em}
  \centering
  \begin{subfigure}[b]{0.95\linewidth}
    \includegraphics[width=\linewidth]{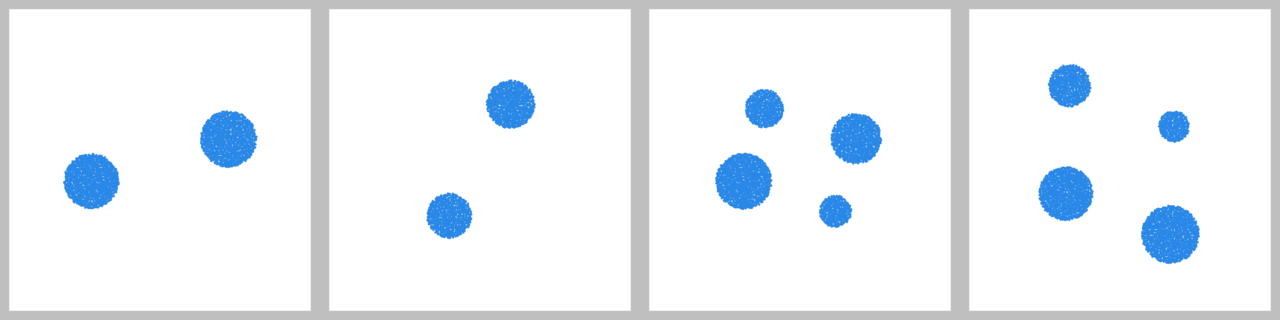}
    \caption*{initial conditions}
    \label{fig:flip2d-data-init}
  \end{subfigure}
  \begin{subfigure}[b]{0.95\linewidth}
    \includegraphics[width=\linewidth]{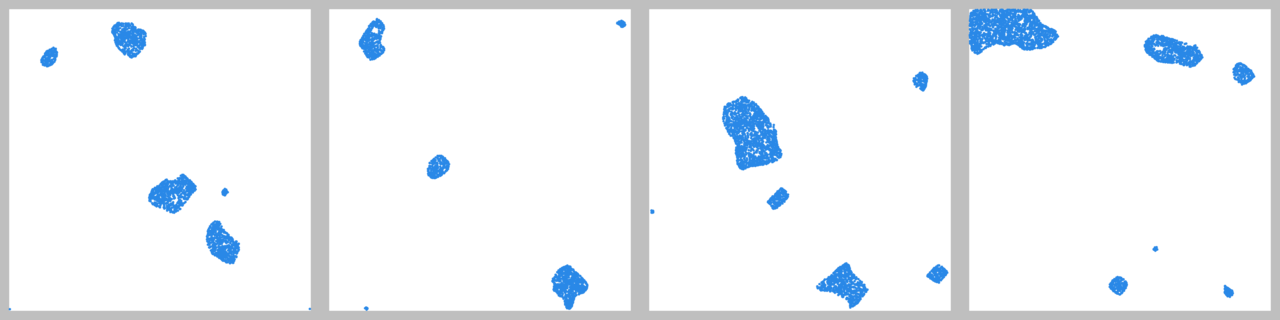}
    \caption{2.5\si{mm}}
    \label{fig:flip2d-data-5mm}
  \end{subfigure}
  \begin{subfigure}[b]{0.95\linewidth}
    \includegraphics[width=\linewidth]{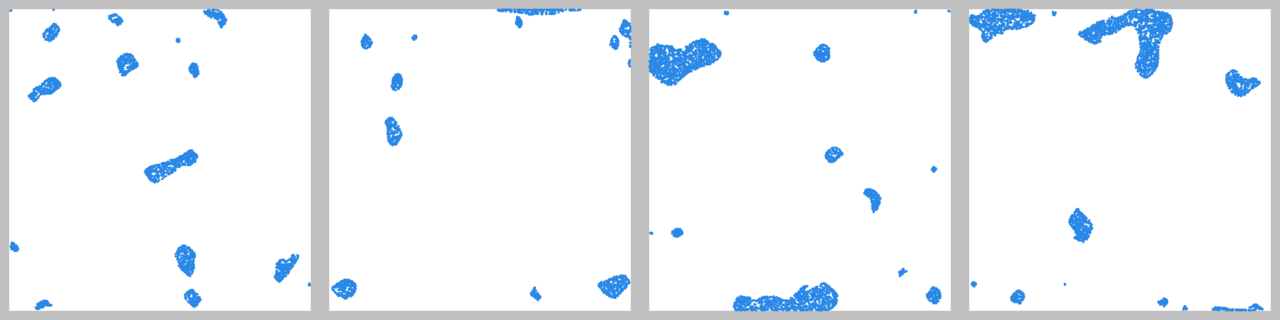}
    \caption{5\si{mm}}
    \label{fig:flip2d-data-1cm}
  \end{subfigure}
  \begin{subfigure}[b]{0.95\linewidth}
    \includegraphics[width=\linewidth]{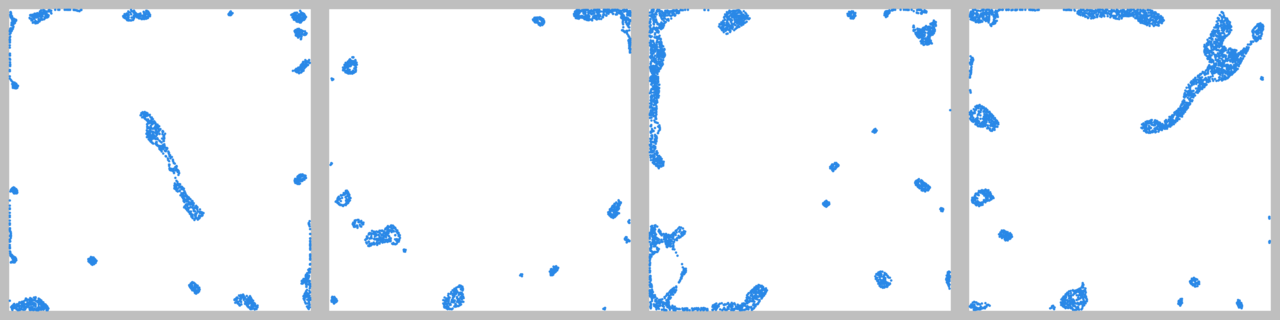}
    \caption{1\si{cm}}
    \label{fig:flip2d-data-2cm}
  \end{subfigure}
  \caption{Example frames of FLIP simulations with surface tension for training
    data generation for three spatial scales: (a) 2.5\si{mm}, (b) 5\si{mm}, and
    (c) 1\si{cm} in grid spacing. The initial conditions for each simulation are
    randomly configured similar to the three-dimensional examples
    (Figure~\ref{fig:training-sim-init}).}
  \label{fig:flip2d-data}
\end{figure}

Our data-driven splash model (i.e., MLFLIP) incites the formation of splashes by
inferring the likelihood and velocity modification of particle. As outlined in
Section~\ref{sec:nns-data}, our approach can employ any available NS solver to
generate the training data. In order to demonstrate that our approach is
agnostic to the choice of solver, we trained our model with two sets of data
that were generated using two different simulation methods: FLIP and SPH. The
FLIP data were generated using randomized initial conditions
(Figure~\ref{fig:flip2d-data}-(\subref{fig:flip2d-data-2cm})) similar to the
three-dimensional example (Figure~\ref{fig:training-sim-init}) with surface
tension effects to incite splashes. On the other hand, SPH often results in a
significant number of splashes due to its direct interparticle interactions. Our
SPH training data were generated using randomized breaking dam simulations as
shown in Figure~\ref{fig:sph-data}. Figure~\ref{fig:mlflip-sph} compares the
results of MLFLIP trained using these two data sets. As shown in the comparison,
both MLFLIP models produce very comparable splashes while demonstrating that our
model successfully learns from both solvers. In addition, we believe that other
NS solvers could likewise be employed to generate training data.

\begin{figure*}[tb]
  \centering
  \includegraphics[width=\linewidth]{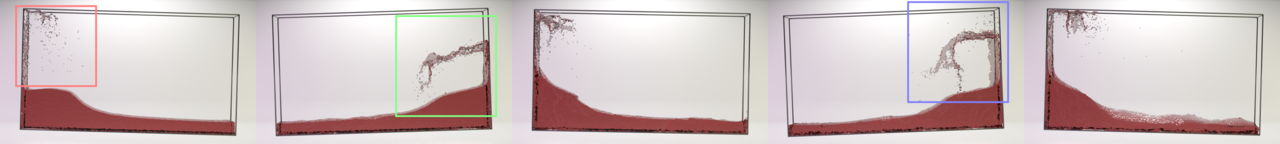}
  \includegraphics[width=0.33\linewidth]{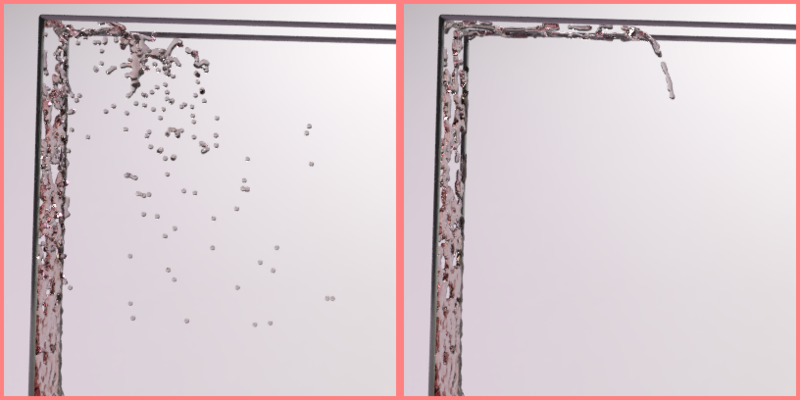}
  \includegraphics[width=0.33\linewidth]{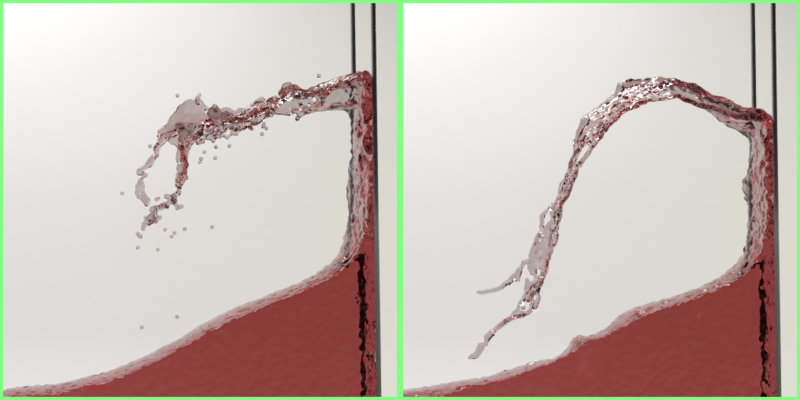}
  \includegraphics[width=0.33\linewidth]{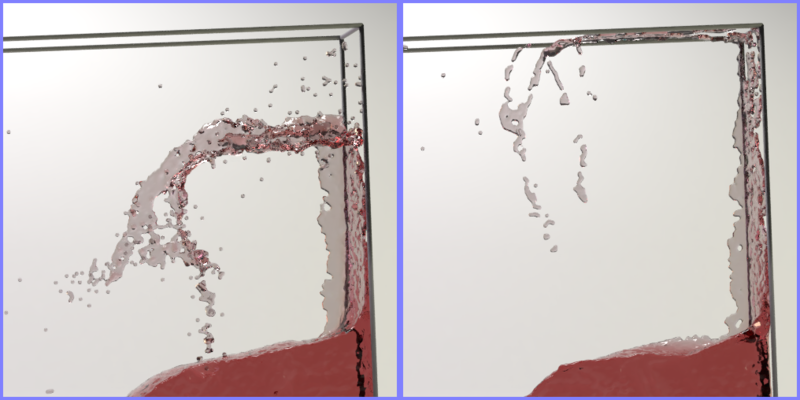}
  \caption{Example frames of wave tank simulations with MLFLIP. The bottom row
    compares (left) MLFLIP with (right) FLIP side-by-side in the selected areas
    of three frames.}
  \label{fig:wave}
\end{figure*}

\subsection{Physical Evaluation}
\label{sec:learning-physics}

Next, we will evaluate the ability of our approach to capture realistic physical
behavior. Due to the turbulent nature of splashing liquids, we rely on a
two-dimensional setup similar to the well known Plateau-Rayleigh instability for
3D flows \cite{gennes2004}. Plateau-Rayleigh instabilities explain the break-up
of tubes of liquid due to surface tension, and it is one of the well known
phenomena leading to droplet formation.

We consider strings of liquid with different initial lengths sampling each with
perturbed FLIP particles. These setups were simulated with surface tension and
viscosity of water with zero gravity. As evaluation metric, we calculate the
number of formed droplets after 1.5 seconds. To generate training data, we
simulated 40 different simulations with a resolution of 1000$\times$100 cells,
which parametrized for a spatial scale of 2.5\si{mm} per cell. As shown in blue
in Figure~\ref{fig:strings}-(\subref{fig:string:ex}), these simulations
accurately resolve the surface tension effects and lead to the droplet counts
shown with blue dots as reference data in
Figure~\ref{fig:strings}-(\subref{fig:string:graph}).

We train our model on this data using a four times smaller base resolution of
250$\times$25, for which we completely omit both surface tension and viscosity.
With this resolution, the string is only ca. 1 cell thick, and hence the surface
tension driven instabilities could not be properly simulated at this scale.
Without surface tension, these simulations would not lead to any droplets
forming. However, when enabling our model, we can accurately simulate the
droplet formation for the liquid string setups. For these tests, we disable the
inference of velocity variance in order to make the results independent of
randomness. The results of our model are shown in orange in
Figure~\ref{fig:strings}-(\subref{fig:string:ex}), and the corresponding droplet
counts can be found in Figure~\ref{fig:strings}-(\subref{fig:string:graph}). As
this graph shows, our model captures the ground truth numbers of droplets very
well across the full range of different lengths. It is also worth pointing out
here that simpler models for generating secondary particles would not be able to
re-create this behavior. As such, this test case successfully demonstrates that
our model learns to represent the underlying physics of droplet formation
faithfully.

\subsection{Secondary Particle Approach}
\label{sec:secondary}

\begin{figure}[tb]
  \centering
  \includegraphics[width=\linewidth]{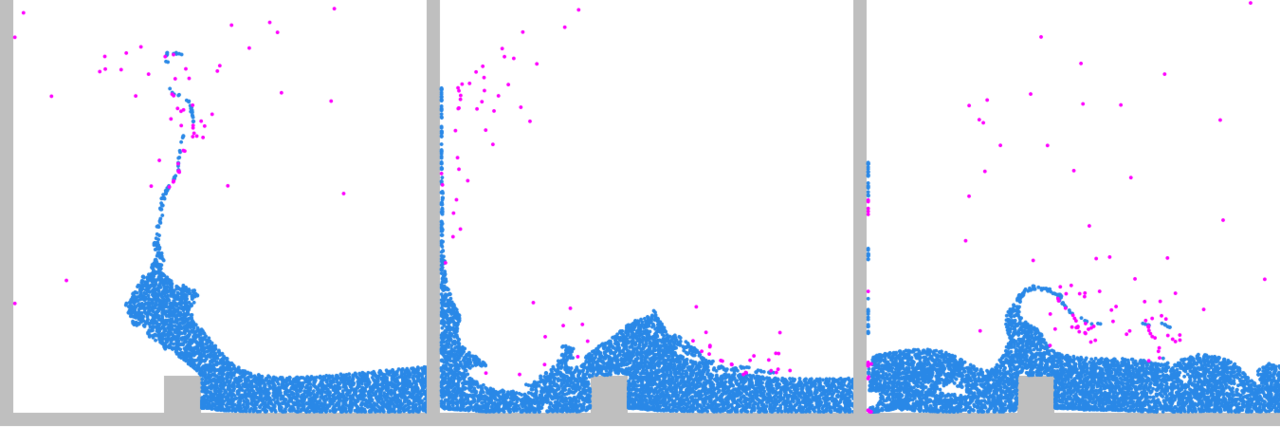}
  \includegraphics[width=\linewidth]{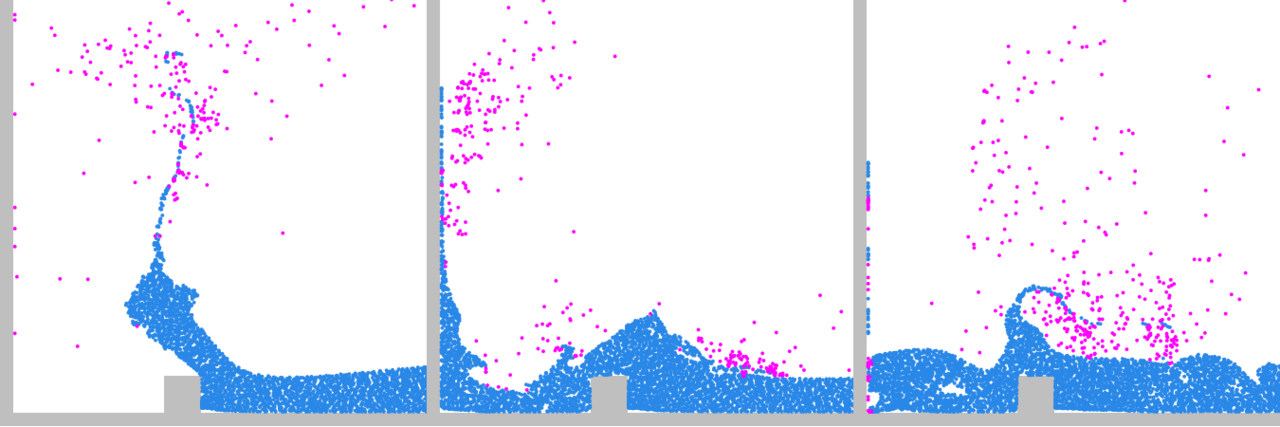}
  \caption{Example frames of MLFLIP with the secondary particle approach. We
    colored secondary particles magenta. Each row uses two different numbers of
    samples (i.e., (top) one and (bottom) four) when we seed secondary particles
    per simulation particle.}
  \label{fig:mlflip-sp}
\end{figure}

While our approach described so far couples particles and bulk motion, we can
modify our algorithm to produce secondary particles. To this end, when our model
classifies simulation particles as splashes, we seed secondary particles at the
same positions as the simulation particles and copy their velocities to the
newly generated secondary particles. After that, the velocity modification is
applied using random numbers. The secondary particle is simulated individually
and removed when it ends up inside of the liquid volume.
Figure~\ref{fig:mlflip-sp} shows the example frames of the secondary particle
model. Here, the splashes were enriched on the pre-simulated FLIP frames. Unlike
the work of Ihmsen \textit{et al.} \shortcite{ihmsen2012}, our model does not
require any manual adjustment of parameters. While secondary splash models from
previous work could potentially produce results similar to ours with enough
manual adjustment of the parameters, the formation of splashes in our model
purely relies on the generated training data. Despite this automated process, we
demonstrate, in the following, that the results can be easily controlled by a
user.

\begin{figure}[tb]
  \centering
  \includegraphics[width=0.95\linewidth]{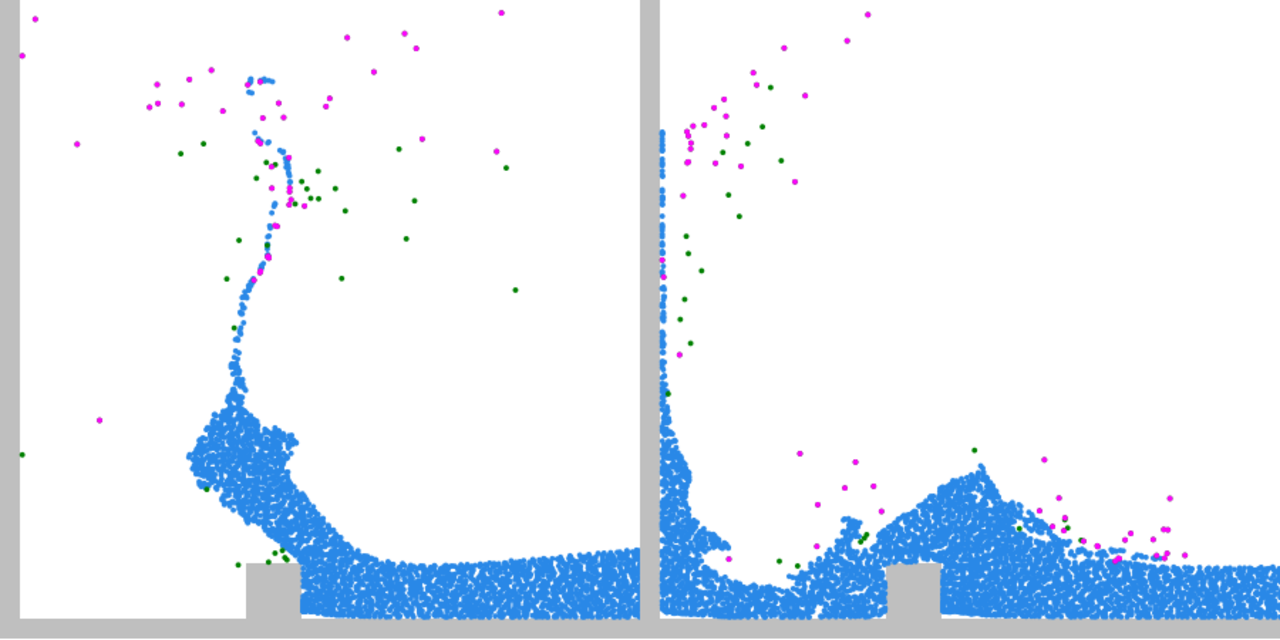}
  \includegraphics[width=0.95\linewidth]{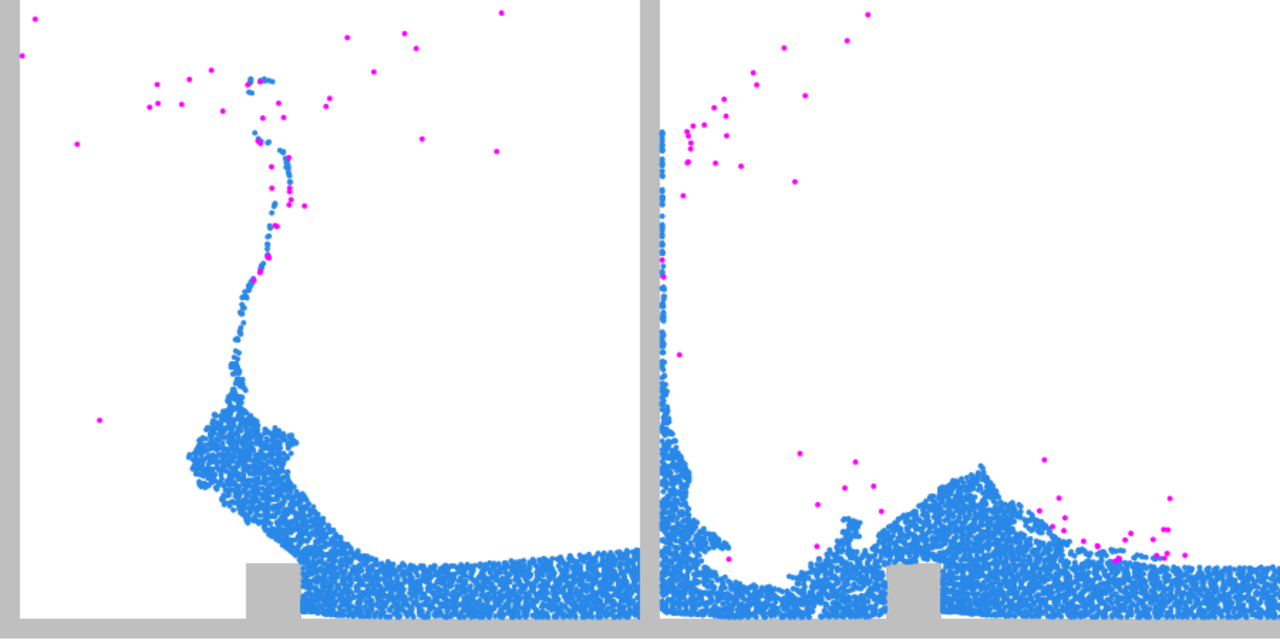}
  \caption{Different numbers of splashes with two control thresholds: (top) 0.2
    and (bottom) 0.8. The difference is highlighted in green at the top; i.e.,
    the green color represents the additional splashes that are generated when
    the lower threshold of 0.2 is used.}
  \label{fig:mlflip-control}
\end{figure}

\begin{figure}[tb]
  \centering
  \includegraphics[width=.49\linewidth]{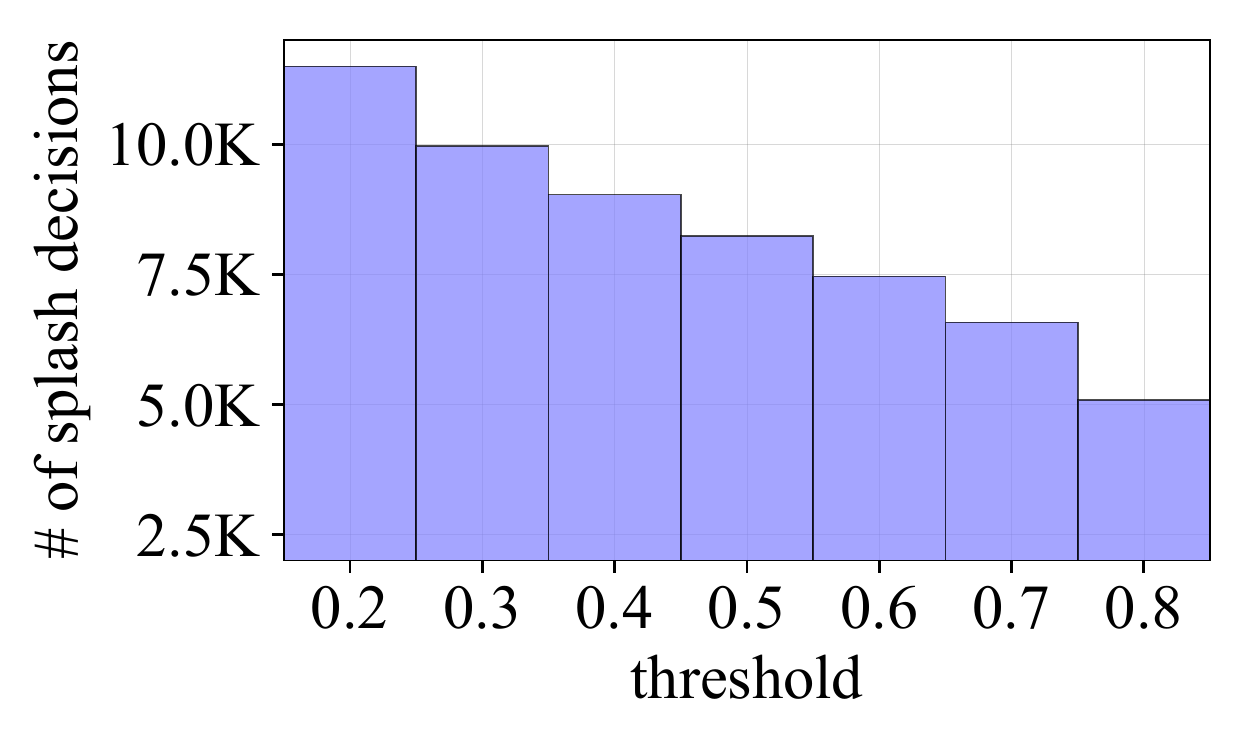}
  \includegraphics[width=.49\linewidth]{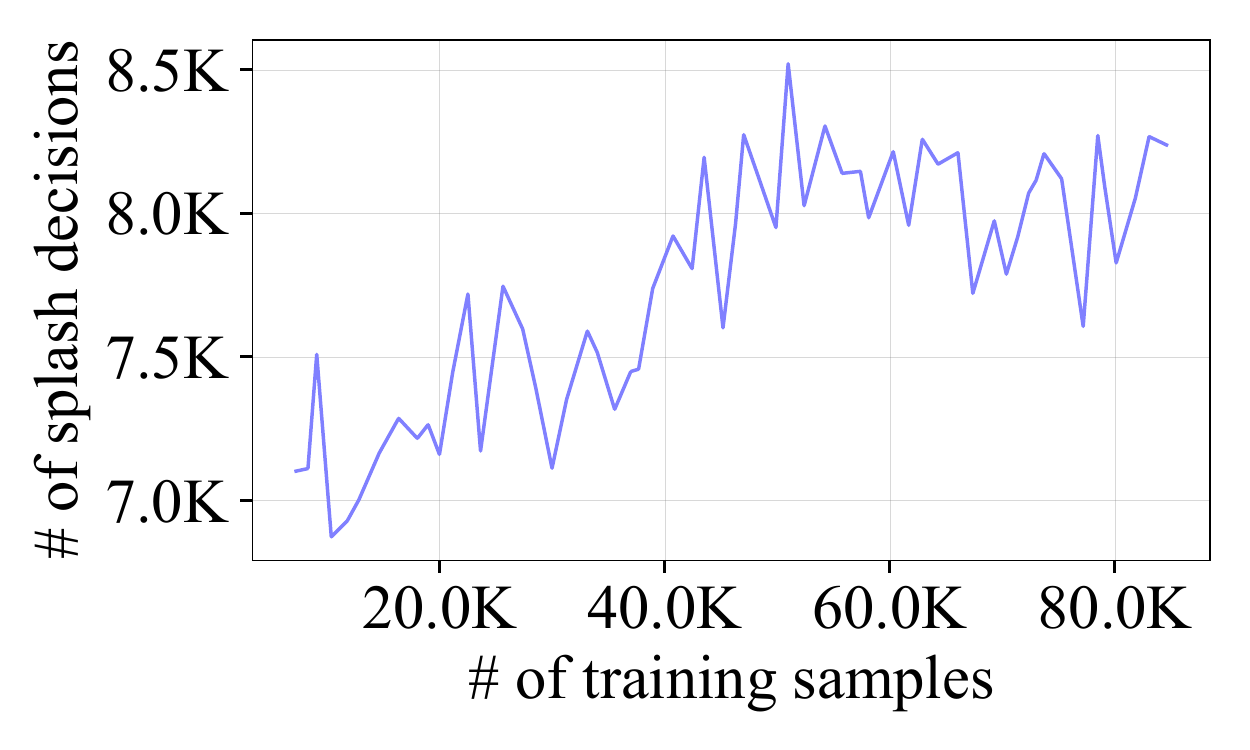}
  \caption{The number of splash decisions with respect to (left) the threshold
    and (right) the number of training samples.}
  \label{fig:cc}
\end{figure}

By virtue of the NN-based classification model, we can readily adopt the output
classification values in order to control the number (i.e., likelihood) of
droplet generation. Rather than classifying splash particles directly via the
output probabilities of $\mathbf{y}_s$, we define a threshold ($\in[0,1]$) and
compare the output probability of forming a splash with this threshold.
Intuitively, a lower threshold classifies the particle with a less likely value
as a splash such that more splashes are generated, and vice versa.
Figure~\ref{fig:mlflip-control} shows the example frames of two thresholds of
0.2 and 0.8. Note that here we adopted the secondary particle approach to assure
the underlying simulations are comparable. The left side of Figure~\ref{fig:cc}
plots the graph of the number of splash decisions with respect to the threshold
value. We found that the number of splashes changes smoothly with respect to the
chosen threshold $\mathbf{y}_s$. Note that positive splash decisions do not
necessarily lead to the generation of visible droplets as some of these
decisions can be reverted as outlined in Section~\ref{sec:nns-flip}.

\begin{figure}[tb]
  \centering
  \includegraphics[width=0.95\linewidth]{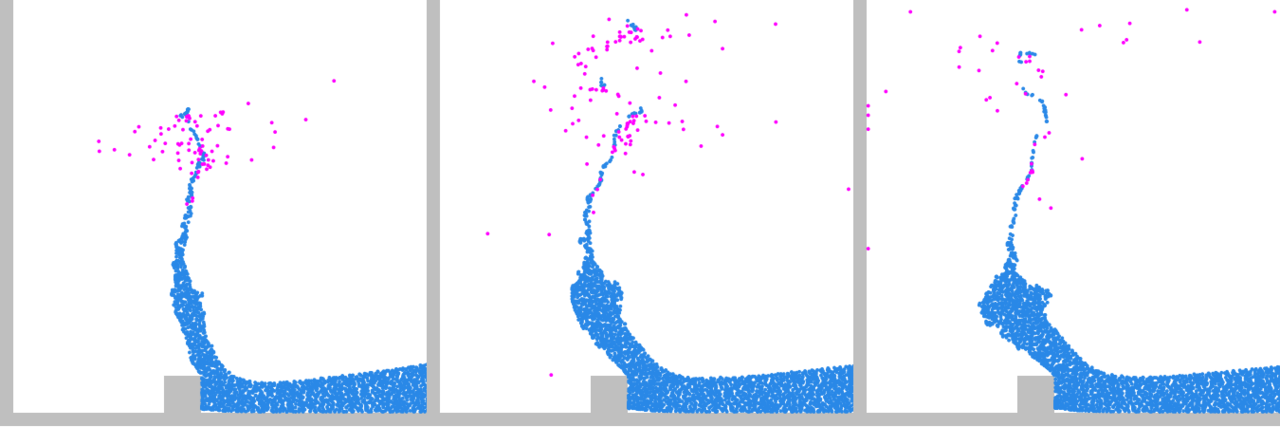}
  \includegraphics[width=0.95\linewidth]{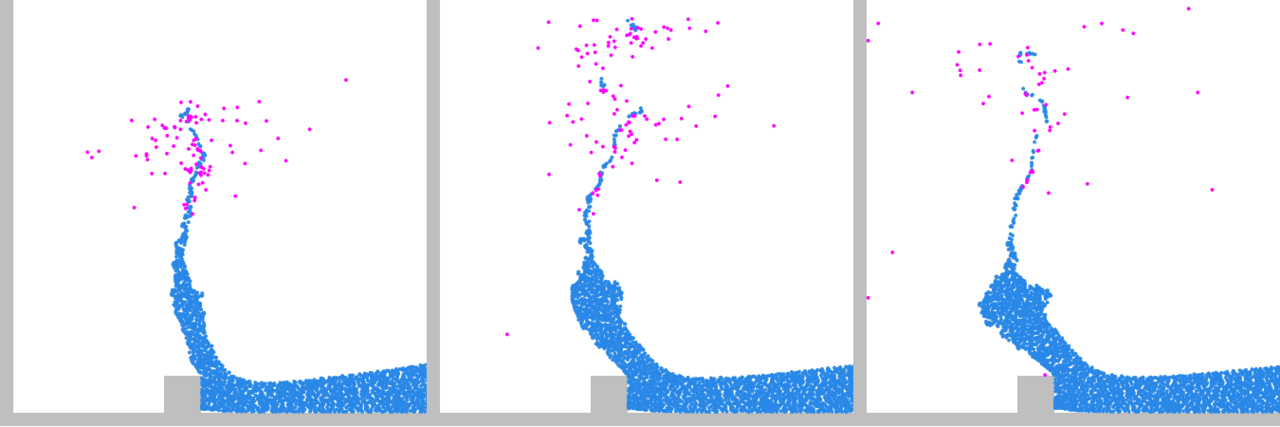}
  \caption{Example frames of MLFLIP in three spatial scales: (left) 5\si{mm},
    (middle) 1\si{cm}, and (right) 2\si{cm} in grid spacing. The top row shows
    the model that is trained using the three scales, whereas the bottom row
    shows the three models that are individually trained for each scale.}
  \label{fig:mlflip-allinone}
\end{figure}

\begin{figure}[t]
  \centering
  \includegraphics[width=0.95\linewidth]{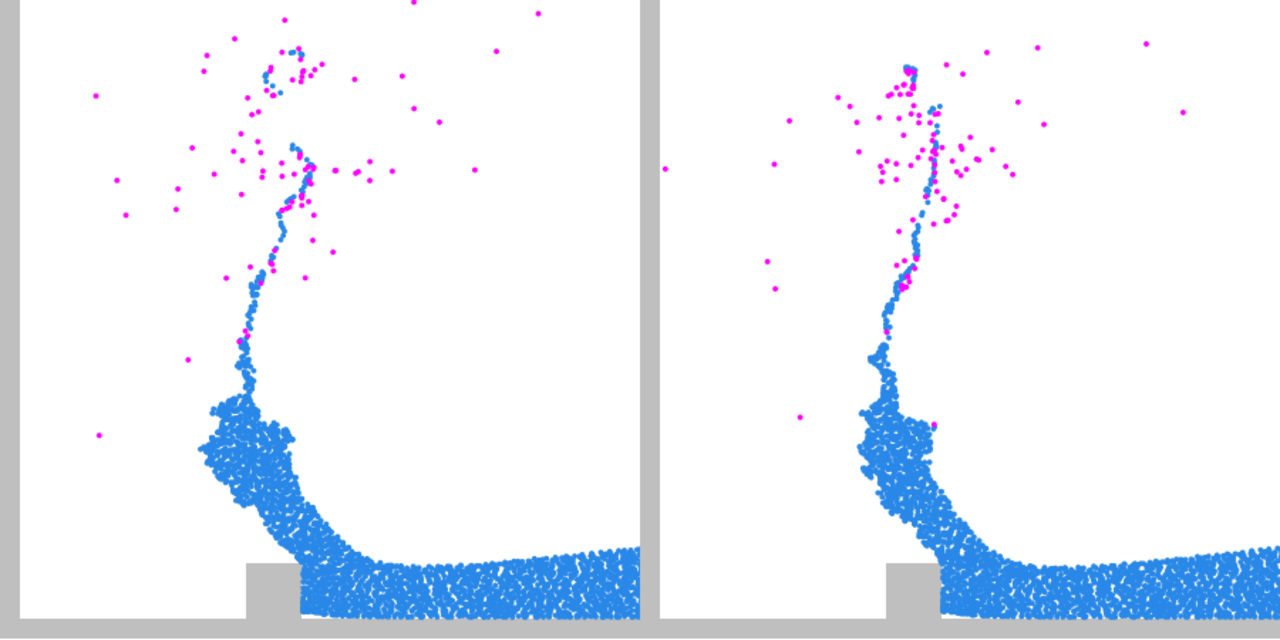}
  \caption{Example frames of the three-scale MLFLIP model from
    Figure~\ref{fig:mlflip-allinone} for two intermediate scales: (left)
    7.5\si{mm} and (right) 1.5\si{cm} that are not part of the training data.}
  \label{fig:mlflip-allinone-m}
\end{figure}

One goal of our approach is to make the model robust to a wide range of spatial
scales. Because of its data-driven nature, the reliable coverage of the model is
often limited to the scale that the training data represent. Rather than
training the model for a certain scale, we extend the model's coverage to a
wider range using \emph{heterogeneous} training data. To this end, we first
generated the data in three scales: 2.5\si{mm}, 5\si{mm}, and 1\si{cm} in grid
spacing (Figure~\ref{fig:flip2d-data}). Then, we used the grid spacing value of
the target resolution as an input feature such that the model can correctly
infer the right output values. Figure~\ref{fig:mlflip-allinone} shows the
results of the extended model for similar surface positions and compares it with
the three models that were individually trained using each set of data for each
scale. We observed that the extended model produces splashes that are comparable
to the three individual models. We also experimented with two intermediate
scales, which are not directly represented by the training data, and found that
the model works robustly as shown in Figure~\ref{fig:mlflip-allinone-m} and is
able to generalize to a wider range of spatial scales. The simulations of
Figure~\ref{fig:mlflip-allinone} and \ref{fig:mlflip-allinone-m} all used the
secondary particle approach.

\myparagraph{Convergence:} In order to investigate if the learning process
converges, we additionally trained our model with different numbers of samples
and monitored the number of splashes generated using the model. Here, we also
adopted the secondary particle approach in order to keep the same conditions for
inference. The right side of Figure~\ref{fig:cc} shows the graph of the number
of splash decisions with respect to the number of training samples. This graph
indicates that the results of our model stabilize for more than ca. 40\si{K}
training samples.

\section{Three-dimensional Results}
\label{sec:results}
This section demonstrates that our model improves the visual fidelity of liquid
simulations in three examples: a \emph{dam}, \emph{stairs}, and a \emph{wave
  tank}. The former two examples represent larger scales than the wave tank
example in length. Hence, we use the 5\si{mm} splash model for both (i.e., dam
and stairs) and 1.5\si{mm} splash model for the wave tank. Figure~\ref{fig:dam}
and \ref{fig:fall} show the comparisons between FLIP and MLFLIP for the dam and
stairs examples. Our model leads to a significant increase in violent and
detailed splashes for this large scale flow. Despite the large number of
splashes, the plausibility of the overall flow is preserved. Likewise, our model
robustly introduces splashes also in the smaller wave tank example as shown in
Figure~\ref{fig:wave}. The smaller real world size in conjunction with the
smaller velocities leads to fewer splashes for this setup.

Our model requires additional calculations for the generation of splashes, and
consequently, this results in a slightly increased runtime. In the dam example,
the average computation time per simulation step was 0.52\si{s} for FLIP, while
it was 0.60\si{s} for MLFLIP. Both simulations used the same grid resolution of
160$\times$150$\times$50, where the grid spacing was 2\si{cm}. In the stairs
example, the runtime was 0.62\si{s} for FLIP and 0.78\si{s} for MLFLIP, and
their grid resolutions were both 100$\times$200$\times$100. In the wave tank
example, the runtime was 0.12\si{s} for FLIP and 0.14\si{s} for MLFLIP. In this
case, the grid resolution was 150$\times$84$\times$10, where the grid spacing
was 0.6\si{cm}.

\begin{figure}[tb]
  \centering
  \includegraphics[width=0.327\linewidth]{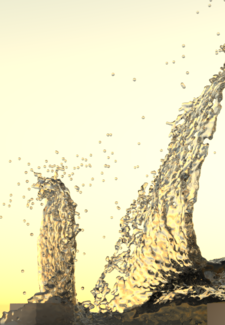}
  \includegraphics[width=0.327\linewidth]{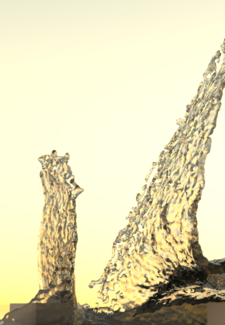}
  \includegraphics[width=0.327\linewidth]{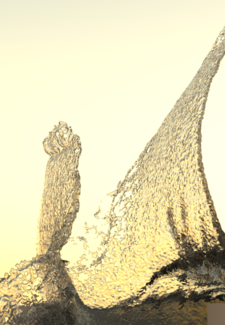}
  \caption{Comparisons among the three simulations: (left) MLFLIP, (middle)
    FLIP, and (right) FLIP with the high resolution.}
  \label{fig:perfcomp}
\end{figure}

We observed that the splashes of our MLFLIP simulation are very difficult to
resolve with regular FLIP simulations even with high resolutions. For the dam
example, Figure~\ref{fig:perfcomp} shows a visual comparison of three
simulations: FLIP and MLFLIP with the same resolution and FLIP with a doubled
resolution of 320$\times$300$\times$100. Despite taking nine times longer per
simulation step (i.e., 5.43\si{s} for FLIP and 0.60\si{s} for MLFLIP), this high
resolution simulation fails to resolve the droplets of our MLFLIP simulation.
Our model successfully generates splash details from a low resolution
simulation, while the high resolution simulation barely improves the number of
splashes despite its high computational cost.

\section{Discussion and Limitations}

Our work shares its goal with the well-known secondary particle approaches. Many
of these approaches have proposed physically inspired heuristics utilizing a
combination of measures, such as curvature, velocity, vorticity, or deformation
gradient to generate secondary particles. This naturally requires a manual
tuning process to find a set of proper parameter values for the different
components. In contrast, our model aims for an automated data-driven algorithm
that does not require any tuning but instead learns a probabilistic model from
reference data. However, with enough manual tuning for each simulation setup,
simpler algorithms for secondary particles could lead to results that are very
similar to the output of our algorithm.
  
Our model directly works with the regular representation of the fluid, i.e., the
simulation particles, and does not require additional helper data structures. In
this way, our model can be easily two-way-coupled with a FLIP simulation
allowing for interactions of the splash effects with the bulk liquid. As a
consequence, our model not only yields improved small scale details but can also
capture realistic droplet formation effects as illustrated in
Figure~\ref{fig:strings}. This effect would be very difficult to re-create with
previous algorithms. Interestingly, our model learns this from a set of examples
instead of requiring a hand crafted analytical expression. Thus, our algorithm
requires additional work for data generation and training but, in this way,
arrives at a more general framework for the data-driven modeling of droplet
effects.

In our data generation step, we typically place several droplets to incite
splashes and randomize the initial velocities of these droplets in a pre-defined
range. With our current model, this limits the inference capabilities, for
instance, when simulations exhibit much larger velocities than the ones present
in the training data. However, it would be possible to detect such cases in
order to trigger generating additional training data. Apart from the data
generation via numerical simulations, we envision that extracting data from real
world experiments \cite{garg2007} will be an interesting alternative for future
work.

Because our model uses the simulation particles to represent droplets, the
details of generated splashes depend on the particle resolution of the
underlying FLIP simulation. This limits the scale of droplets to the size of a
single simulation particle. Our model could be extended to a scale variant
droplet model by adopting an adaptive approach \cite{ando2013} and learning the
formation of different scale droplets from physical parameters. The Ohnesorge
number \cite{lefebvre1988} to characterize the dispersion behavior of liquids
would be a good candidate here.

Although the splashes incited by our model transfer momentum when they merge
with the liquid volume, our model currently does not compute interactions among
droplets \cite{jones2017}. Finally, we only experimented with the formation of
splash droplets. However, we envision that our model could be extended to other
complex sub-grid effects such as bubbles, capillary waves, and foam, which are
highly expensive to compute with regular free-surface liquid simulations. We
believe that droplet interactions and additional small-scale phenomena are
important for believable simulations; thus, it will be very interesting to
explore how our NN-based model can extend to these directions in future work.

\section{Conclusions}

This paper introduced a new data-driven splash generation model that utilizes
machine learning techniques with NNs. We demonstrated that our models
successfully learn the formation of splashes at different physical scales from
the training data with the training process. Our model leads to improved
splashing liquids and successfully learns to recognize the relevant mechanisms
for droplet formation from the pre-computed data. Moreover, we demonstrated an
extension of our model for secondary particles, and we evaluated the learned
models with a variety of complex tests. Overall, our experiments highlight that
our approach provides a very good basis for learning a wide range of realistic
physical effects.

\section*{Acknowledgments}

This work is supported by the ERC Starting Grant 637014.

\bibliographystyle{eg-alpha-doi}

\bibliography{references}

\end{document}